\documentclass[superscriptaddress,twocolumn,showpacs,aps,prb,longbibliography]{revtex4}
\usepackage{amsmath,amssymb,bm,mathtools}
\usepackage{amsfonts}
\usepackage{graphicx}
\usepackage{dcolumn}
\usepackage{natbib,hyperref}
\usepackage{bm}
\usepackage{soul}
\usepackage{epstopdf}
\usepackage{color}
\usepackage{comment}
\usepackage{array}

\graphicspath{{dmchainfigures/}{figures/}} % Location of the graphics files

\begin{document}

\title{Ising order in a magnetized Heisenberg chain subject to a uniform Dzyaloshinskii-Moriya interaction}
\author{Yang-Hao Chan}
\affiliation{Institute of Atomic and Molecular Sciences, Academia Sinica, Taipei 10617, Taiwan}

\author{Wen Jin}
\affiliation{Department of Physics and Astronomy, University of Utah, Salt Lake City, Utah 84112, USA}

\author{ Hong-Chen Jiang}
\affiliation{Stanford Institute for Materials and Energy Sciences, SLAC National Accelerator
Laboratory and Stanford University, 2575 Sand Hill Road, Menlo Park, California 94025, USA}

\author{Oleg A. Starykh}
\affiliation{Department of Physics and Astronomy, University of Utah, Salt Lake City, Utah 84112, USA}

\date{\normalsize\today}
\begin{abstract}
We report a combined analytical and density matrix renormalized group study of the antiferromagnetic $XXZ$ spin-1/2 Heisenberg chain subject to a uniform Dzyaloshinskii-Moriya (DM) 
interaction and a transverse magnetic field. The numerically determined phase diagram of this model, which features two ordered Ising phases and a critical Luttinger liquid one 
with fully broken spin-rotational symmetry, agrees well with the predictions of Garate and Affleck [Phys. Rev. B $\mathbf{81}$, 144419 (2010)].
We also confirm the prevalence of the $N^z$ N\'eel Ising order in the regime of comparable DM and magnetic field magnitudes.
\end{abstract}

\maketitle

%\tableofcontents

\section{Introduction}

The physics of quantum spins is at the center of modern condensed matter research. The ever present spin-orbit interactions, long considered to be an unfortunate annoying feature of real-world materials, 
are now recognized as the key ingredient of numerous spintronics applications \cite{Alicea2012,Manchon2015} and the crucial tool for constructing topological phases \cite{Kitaev2001,Kitaev2006}.

In magnetic insulators atomic spin-orbit coupling leads, via superexchange mechanism, to an asymmetric spin exchange ${\bf D}_{ij} \cdot {\bf S}_i \times {\bf S}_j$, 
known as Dzyaloshinskii-Moriya (DM) interaction \cite{Dzyaloshinskii1958,Moriya1960}, between localized spins ${\bf S}$ at sites $i$ and $j$. 
Classically, such an interaction induces incommensurate spiral correlations in the plane perpendicular to the DM vector ${\bf D}_{ij}$. Incommensurability of the spin spiral is determined by $D/J$, 
where $J$ is the magnitude of the isotropic exchange interaction between nearest spins. This ratio is typically quite small, resulting in spiral correlations with very long wavelengths. 
It was realized long ago that the external magnetic field, applied {\em perpendicular} to the DM axis, causes strong modification of the spiral state and produces a chiral soliton lattice, 
a periodic array of domains, commensurate with the lattice, separated by $2\pi$-domain walls (solitons) \cite{Dzyaloshinskii1965}. This incommensurate structure undergoes a continuous incommensurate-commensurate
transition into a uniform ordered state at  a rather small critical magnetic field of the order of $D$. \cite{Dzyaloshinskii1965,Zheludev1998,Togawa2012} Such potential tunability makes this interesting class of magnetically-ordered materials particularly attractive for multiferroics and spintronics applications \cite{Cheong2007,Kishine2015}.

It is not well understood how strong quantum fluctuations modify this classical picture. 
To this end, and also having in mind several spin-1/2 quasi-one-dimensional quantum magnets \cite{Povarov2011,Halg2014,Smirnov2015} for which this consideration is highly relevant, we 
investigate here the joint effect of a uniform DM interaction $D {\hat z} \cdot {\bf S}_i \times {\bf S}_{i+1}$ and a 
transverse magnetic field $h S^x_i$ on the low-energy properties of the antiferromagnetic spin-1/2 Heisenberg chain with a weak $XXZ$ anisotropy $\Delta$.
Our goal is to quantitatively check, with the help of the state-of-the-art density-matrix renormalization group (DMRG) calculation, predictions of the recent field-theoretic studies of this interesting 
problem \cite{Gangadharaiah2008,Garate2010,Sun2015}. Garate and Affleck,\onlinecite{Garate2010}, found that quantum fluctuations destroy the chiral soliton lattice and replace 
it with a critical Luttinger-liquid (LL) phase.
Additionally, the model is found to support two distinct ordered phases with staggered Ising order along directions perpendicular to the external field ${\bf h}$. Regions of stability of these Ising phases are 
found to differ significantly from the classical expectations \cite{Gangadharaiah2008,Garate2010}. In particular, when the magnitudes of the DM interaction $D$ and magnetic field $h$ are comparable to each other,
the Ising-like longitudinal spin-density wave order (of $N^z$ kind; see below) is found to extend deep into the classically forbidden $\Delta \leq 1$ region.

The outline of the paper is as follows. Section~\ref{sec:theory} reviews the field-theoretic arguments and Sec.~\ref{sec:theory-phase} summarizes the quantum phase diagram. The main DMRG results are presented in 
Sec~\ref{Sect:Numerics}. Section~\ref{Sect:FiniteSizeEffect} focuses on understanding the strong finite-size effects observed in our study. Numerous Appendixes provide technical details of our analytical 
(Appendixes \ref{app:bosonization}-\ref{app:orderp}) and numerical (Appendix \ref{app:DMRG}) calculations.

%%%%%%%%%%%%%%%%%%%%%%%%%%%%%%%%%%%%%%%%%%%
%%%%%Analytical Part starts here%%%%%%%%%%%%%%%%%%%%%%
%%%%%%%%%%%%%%%%%%%%%%%%%%%%%%%%%%%%%%%%%%%

%\part{Analytical Part}\label{part2}
\section{Hamiltonian of the model}
\label{sec:theory}

We consider antiferromagnetic Heisenberg spin-$1/2$ chains subject to a uniform DM interaction and a transverse external magnetic field. 
The system is described by the Hamiltonian
\begin{equation}
\begin{split}
{\cal H}&=J\sum_{i} \left[{S}^{x}_{i} {S}^{x}_{i+1}+{S}^{y}_{i} {S}^{y}_{i+1}+\Delta {S}^{z}_{i}{S}_{i+1}^{z}\right]\\
&\qquad -\sum_{i} D \hat{z}\cdot({\bm S}_{i}\times {\bm S}_{i+1})-{h}\sum_{i}{S}_{i}^x,\\
\end{split}
\label{eq:H_0}
\end{equation}
where ${\bf S}_{i}$ is the spin-$1/2$ operator at site $i$, $J$ denotes antiferromagnetic exchange coupling between nearest neighbors, and 
$\Delta \approx 1$ parametrizes small Ising anisotropy. 
The DM interaction is parametrized by the DM vector ${\bf D}=D\hat{z}$, which is uniform along the chain. 
We consider $D/J \ll 1$, which is the most natural limit relevant for real materials\cite{Povarov2011,Halg2014,Smirnov2015, Dmitrienko2014}. 
In addition to twisting spins around the ${\bm D}$ axis, the 
uniform DM interaction slightly renormalizes Ising anisotropy \cite{Garate2010} by an amount proportional to $D^2/J^2$.
Here $h$ denotes the strength of the applied transverse magnetic field.

\subsection{Hamiltonian in the low-energy limit}
\label{subsec:low-energy limit}
In the low-energy continuum limit, the bosonized Hamiltonian of the problem reads \cite{Gangadharaiah2008,Schnyder2008,Garate2010}
\begin{equation}
{\cal  H}_{\rm chain}=\tilde{H}_0+\tilde{H}_{\rm bs},
\end{equation}
where $\tilde{H}_0$ has a quadratic form in terms of Abelian bosonic fields $(\varphi, \vartheta)$ (see Appendix~\ref{app:bosonization} for details) and
the Zeeman and DM interaction terms [second line in Eq.~\eqref{eq:H_0}] are absorbed in $\tilde{\cal H}_0$ by a chiral rotation and subsequent linear shift of field $\varphi$ 
as described in Appendix~\ref{app:chiral rotation}. 
The harmonic Hamiltonian $\tilde{H}_0$ is perturbed by the 
chain backscattering $\tilde{H}_{\rm bs}$ describing the residual backscattering interaction between right- and left-moving spin modes of the chain. It consists of several contributions \cite{Garate2010,Gangadharaiah2008,Jin2017}
\begin{equation}
\begin{split}
\tilde{H}_{\rm bs}&= H_A+H_B+H_C+H_{\sigma}, \\
H_{A}&=\pi v y_A\int \mathrm{d}x(M_{R}^{z}M_{L}^{+}e^{it_{\varphi}x}-M_{R}^{+}M_{L}^{z}e^{-it_{\varphi}x}+{\rm H.c.}),\\
H_{B}&=\pi v y_B\int \mathrm{d}x(M_{R}^{+}M_{L}^{-}e^{-i2t_{\varphi}x}+{\rm H.c.}), \\
H_{C}&=\pi v y_C\int \mathrm{d}x (M_{R}^{+}M_{L}^{+}+{\rm H.c.}), \\
H_{\sigma}&=-2\pi v y_{\sigma}\int \mathrm{d}x M_{R}^{z}M_{L}^{z}.
\end{split}
\label{eq:Hbs}
\end{equation}
Here ${\bf M}_{L}(x)$ and ${\bf M}_{R}(x)$ are the uniform left- and right-moving spin current operators defined in Appendix~\ref{app:chiral rotation}, and we
use the following notations
\begin{equation}
\label{eq:yC}
\begin{gathered}
y_C\equiv\frac{1}{2}(y_x-y_y), \;\;y_B\equiv\frac{1}{2}(y_x+y_y),\;\; y_{\sigma}\equiv-y_z,\\
t_{\varphi}\equiv\frac{\sqrt{D^2+h^2}}{v}.
\end{gathered}
\end{equation}
Initial values of the coupling constants are given by\cite{Garate2010,Jin2017}
\begin{equation}
\begin{split}
y_x(0)&=-\frac{g_{\rm bs}}{2\pi v}[(1+\frac{\lambda}{2})\cos\theta^- +\frac{\lambda}{2}],\\
y_y(0)&=-\frac{g_{\rm bs}}{2\pi v},\\
y_z(0)&=-\frac{g_{\rm bs}}{2\pi v}[(1+\frac{\lambda}{2})\cos\theta^- -\frac{\lambda}{2}],\\
y_A(0)&=\frac{g_{\rm bs}}{2\pi v}(1+\frac{\lambda}{2})\sin\theta^-,\qquad
\end{split}
\label{initial}
\end{equation}
where the magnitude of backscattering $g_{\rm bs} \approx 0.23 \times (2\pi v)$ (see Ref. \onlinecite{Garate2010} for details),
\begin{equation}
\begin{gathered}
\theta^-=2\theta_0,\quad \theta_0= -\arctan(D/h),
\end{gathered}
\end{equation}
and $v\simeq J\pi a/2$ is the spin velocity, with $a$ the lattice constant. 
%{\color{red} The parameter $\beta=2\pi R$ is related to the ``compactification radius" $R$ in the sine-Gordon (SG) model. 
%In the absence of the external field, the SU(2) invariant Heisenberg chain is characterized by $2\pi R^2=1$. - REMOVE!}

The $XXZ$ anisotropy is parametrized by $\lambda$ \cite{Garate2010},
\begin{equation}
\lambda=c(1-\Delta+\frac{D^2}{2J^2}).
\label{eq:lambda}
\end{equation}
The constant $c=(4v/g_{\rm bs})^2$ is about $7.66$ from the Bethe-ansatz solution [see (B2) in Ref. \onlinecite{Garate2010}]. 
The oscillating factor $e^{it_{\varphi}x}$ in \eqref{eq:Hbs} is introduced by the effective transverse field $h_{\rm eff}=\sqrt{h^2+D^2}$,
which accounts for the combined effect of the magnetic field and DM interaction [see \eqref{eq:effV}].

Our task is to identify the most relevant coupling in perturbation~\eqref{eq:Hbs}, which is accomplished by the renormalization group (RG) analysis. 
%--figure:KT flow 
\begin{figure}[!]
	\centering
	\includegraphics[width=0.35\textwidth]{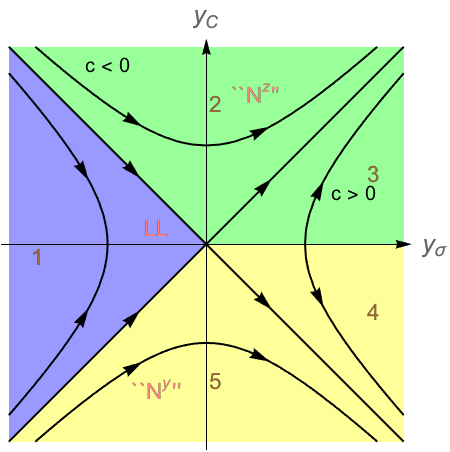}
	\caption{(Color online) Solution of Kosterlitz-Thouless equations \eqref{yyc}. Different symbols and colors depict  Ising phases N$^z$ (green region) and N$^y$ (yellow region)  
	and the critical Luttinger liquid phase (purple region) according to the RG flow criteria summarized in Table~\ref{table:c}. }
	\label{fig:kt}
\end{figure}
\newcolumntype{K}[1]{>{\centering\arraybackslash}p{#1}}
\begin{table}[!]
	%\centering
	\begin{center}
		{\renewcommand{\arraystretch}{1.2}%
			\begin{tabular}{ K{1cm}|  K{1cm}| K{1cm}| K{1cm}| K{1cm}| K{1cm}}
				\hline\hline
				\centering
				Region & 1 &  2 & 3 & 4  & 5 \\ 
				\hline
				$y_{C}(0)$ & $+/-$ & $\;\;+\;\;$ & $\;\;+\;\;$ & $\;\;-\;\;$ & $\quad-\quad$ \\
				\hline
				$y_{\sigma}(0)$ & $-$ & $-$/$+$ & $+$ & $+$ & $-$/$+$ \\
				\hline
				$C$ & $+$ & $-$ & $+$ & $+$ & $-$ \\
				\hline
				$y_C(\ell^*)$ & 0 & $\;+\infty\;$ & $\;+\infty\;$ &$\;-\infty\;$ &$\;-\infty\;$\\
				\hline
				$y_\sigma(\ell^*)$ & finite & $\;+\infty\;$ & $\;+\infty\;$ &$\;+\infty\;$ &$\;+\infty\;$\\
				\hline
				$y_B(\ell^*)$ & finite & finite & finite & finite & finite\\
				\hline
				State & LL & ``$N^z$" & ``$N^z$" & ``$N^y$" & ``$N^y$"\\
				\hline\hline
			\end{tabular}}
			\caption{Signs and values of $y_C$, $y_{\sigma}, and C$ corresponding the KT flow in Fig.~\ref{fig:kt}. Here $\ell^*$ is the critical RG scale at which one
				(or several) coupling constant reaches the strong-coupling limit (and becomes of order one).}
			\label{table:c}
		\end{center}
	\end{table}
	
\subsection{Two-stage RG}
\label{subsec:2steprg}

Renormalization group equations for coupling constants of the backscattering interaction \eqref{eq:Hbs} are obtained with the help of the operator product expansion \cite{fradkin,ope} technique and read
\begin{equation}
\begin{split}
\frac{dy_x}{dl}&=y_yy_z,\quad
\frac{dy_y}{dl}=y_x y_z+y_A^2,\\
\frac{dy_z}{dl}&=y_x y_y,\quad
\frac{dy_A}{dl}=y_yy_A.\\
\end{split}
\label{eq:RG_bs}
\end{equation}
The presence of oscillating $e^{i t_\varphi x}$ factors implies the appearance of the spatial scale, proportional to $1/t_\varphi$, and, correspondingly,
of the RG scale $\ell_{\varphi}$
\begin{equation}
\begin{gathered}
\ell_{\varphi}=\ln(\frac{1}{a_0t_\varphi}) = \ln[\frac{1}{20.4}\frac{\pi}{2}\frac{1}{\sqrt{D^2+h^2}}].
\label{eq:lphi}
\end{gathered}
\end{equation}
where, $a_0=20.4a$ is the ultraviolet RG cutoff length scale \cite{Garate2010} (see Ref.~\onlinecite{Garate2010}
for details of how the choice of the initial value for $g_{\rm bs}$ also determines $a_0$). 

For $\ell < \ell_{\varphi}$ oscillations due to $e^{it_{\varphi}x}$ can be neglected and the full set of RG equations \eqref{eq:RG_bs} has to be solved numerically.
Once RG time $\ell>\ell_{\varphi}$, strong oscillations in $H_A$ and $H_B$ result in the disappearance of these terms from the Hamiltonian.
Correspondingly, we can set $y_A(\ell)=0$ and $y_B(\ell)=0$ in the RG equations. Therefore, at this second stage, the RG equations simplify to [see Eq.~\eqref{eq:yC}]
\begin{equation}
\frac{dy_C}{d\ell}=y_Cy_\sigma,\;\;\;
\frac{dy_\sigma}{d\ell}=y_C^2.\\
\label{eq:RG_tot2}
\end{equation}
These are the well-known Kosterlitz-Thouless (KT) equations, the  analytic solution of which is summarized in Appendix~\ref{app:ktsolution}.
The initial values of backscattering couplings at the second stage are
\begin{equation}
\begin{split}
y_C(\ell_\varphi)&=(y_x(\ell_{\varphi}) - y_y(\ell_{\varphi}))/2 \\
&\to -\frac{g_{\rm bs}}{4\pi v}[(1+\frac{\lambda}{2})\cos\theta^{-}-1+\frac{\lambda}{2}],\\
y_{\sigma}(\ell_\varphi)&=-y_z(\ell_{\varphi}) \to \frac{g_{\rm bs}}{2\pi v}[(1+\frac{\lambda}{2})\cos\theta^{-}-\frac{\lambda}{2}],\\
C&=y_{\sigma}(\ell_\varphi)^2-y_C(\ell_\varphi)^2,\\
\end{split}
\label{yyc}
\end{equation}
where $\cos\theta^{-}=(h^2-D^2)/(h^2+D^2)$ and $C$ is the constant of motion, with $dC/d\ell =0$. Expressions following the right-arrow sign $\rightarrow$ in the above equations
pertain to the situation when the first stage of RG flow, $\ell < \ell_{\varphi}$, can be skipped. This is the case of strongly oscillating $e^{i t_\varphi x}$ factors in Eq.~\eqref{eq:Hbs}, 
when all the oscillating terms in the backscattering Hamiltonian can be omitted from the outset and, correspondingly, $y_a(\ell_{\varphi}) \approx y_a(0)$.
Formally, this limit corresponds to a {\em negative} $\ell_\varphi$ as defined in Eq.~\eqref{eq:lphi}.

\subsection{Ising orders}

We have identified five distinct regions with different signs of $y_{C,\sigma}$ and
integration constant $C$, which result in different RG flows. The boundaries of these regions depend on the initial values of $y$'s and $C$.
When the first-stage flow can be skipped, which happens for sufficiently large $h_{\rm eff}$ such that formally 
$\ell_\varphi < 0$, as discussed at the end of Sec.~\ref{subsec:2steprg},
then the dependence on initial values can be directly translated into that on $h/D$ ($\cos\theta^-$) and $\lambda$ ($\Delta$ and $D/J$).
These results are summarized in Table~\ref{table:c} and Fig.~\ref{fig:kt}, which shows what orders are promoted in different regions.  

Small $t_\varphi$ results in $\ell_\varphi > 0$ and a two-step RG analysis is required, as explained above. 
Once the RG equations \eqref{eq:RG_bs} are integrated to $\ell = \ell_\varphi$, all the oscillating terms must be dropped and only two momentum-conserving terms, 
$H_c$ and $H_{\sigma}$, remain present in the Hamiltonian.

In terms of Abelian fields $(\varphi,\vartheta)$, the interaction $H_C$ is nonlinear, $H_C\propto y_C \cos[2\sqrt{2\pi}\vartheta]= y_C \cos[2\beta\vartheta]$,
while $H_\sigma \propto (\partial_x \varphi)^2 - (\partial_x \vartheta)^2$ and describes renormalization of $\beta$ (see Appendix \ref{subsec:hcandhs}).
(We neglect marginal renormalization of the spinon's velocity $v \to v \sqrt{1-y_\sigma^2/2}$.)
The ground state of the chain is determined by the ordering of the $\vartheta$ field.

It is important to understand how the chiral rotation, which led to \eqref{eq:Hbs}, affects staggered magnetization and dimerization. Arguments in 
Appendix~\ref{app:chiral rotation} show that staggered magnetization ${\bf N}$ and dimerization $\epsilon$ in the laboratory frame are related
to those in the rotated frame, ${\bm {\mathcal N}}$ and $\xi$, as follows:
\begin{eqnarray}
{\bm N} &=& (-{\cal N}^{z}, \cos\theta_0 {\cal N}^{y} + \sin\theta_0 \xi, {\cal N}^{x}),\nonumber\\
 \epsilon &=& \cos\theta_0 \xi - \sin\theta_0 {\cal N}^{y}.
 \label{eq:Neps}
 \end{eqnarray}
Further, a shift of the $\varphi$ field by $t_\varphi x$ [Eq.~\eqref{eq:shift}] introduces a $t_\varphi x$ dependence in the arguments of fields ${\cal N}^{z}$ and 
$\xi$ [Eq.~\eqref{shift:J-N}], but does not affect the ${\cal N}^{x,y}$ pair.

Flow of the KT equations \eqref{eq:RG_tot2}  to strong coupling implies development of the expectation value for the $\vartheta$ field.
When $y_C\to +\infty$ for $\ell \to \infty$, the energy is minimized by $\sqrt{2\pi}\vartheta=(2k_1+1)\pi/2$, with $k_1$ an integer, and ${\cal N}^x\propto -\sin\sqrt{2\pi}\vartheta\neq 0$.
This means that in the original frame there is an Ising order $N^z\neq 0$, and following Ref.~\onlinecite{Garate2010} we name this state ``$N^z$". 
The long-range order (staggered magnetization) in the laboratory frame is {\em commensurate},
\begin{equation}
\langle \bm N(x) \rangle \propto \langle\sin(\sqrt{2\pi}\vartheta)\rangle {\bf z} \propto (-1)^{k_1+1}  {\bf z}.
\label{eq:order1}
\end{equation}
In the case of $y_C\to -\infty$ the energy is minimized by $\sqrt{2\pi}\vartheta=k_2\pi$, with $k_2$ an integer, and ${\cal N}^y\propto\cos\sqrt{2\pi}\vartheta\neq 0$.
Therefore, the Ising order is now along the ${\bf y}$ axis, $N^y\neq 0$, and we name it $N^y$. In addition, according to Eq.~\eqref{eq:Neps},
the finite expectation value of ${\cal N}^y$ implies finite staggered magnetization $\epsilon$. \cite{Garate2010}
Therefore, the $N^y$ phase is characterized by the coexistence of commensurate Ising N\'eel and dimerization orders
\begin{equation}
\begin{split}
\langle \bm N(x) \rangle &\propto \cos\theta_0 \langle \cos(\sqrt{2\pi}\vartheta)\rangle {\bf y} \propto \cos\theta_0 (-1)^{k_2} {\bf y}, \\
\epsilon &\propto - \sin\theta_0 \langle \cos(\sqrt{2\pi}\vartheta)\rangle \propto \sin\theta_0(-1)^{k_2+1}.
\end{split}
\label{eq:order2}
\end{equation}

Finally, a gapless regime of $y_C \to 0$ for $\ell \to \infty$ is also possible \cite{Garate2010}. Here the Hamiltonian is purely quadratic and describes a
critical LL phase with algebraic correlations even though the spin rotational symmetry is fully broken \cite{Garate2010,Giamarchi1988};
(see Appendix~\ref{app:LL} for detailed arguments).
As described in the Introduction, the LL state is the quantum version of the classical chiral soliton lattice phase. This is a critical state
with {\em incommensurate} (and anisotropic) spin correlations which decay algebraically with distance.

\section{Phase diagram of the quantum model}
\label{sec:theory-phase}

\begin{figure}[t]
	\centering
	\includegraphics[width=0.35\textwidth]{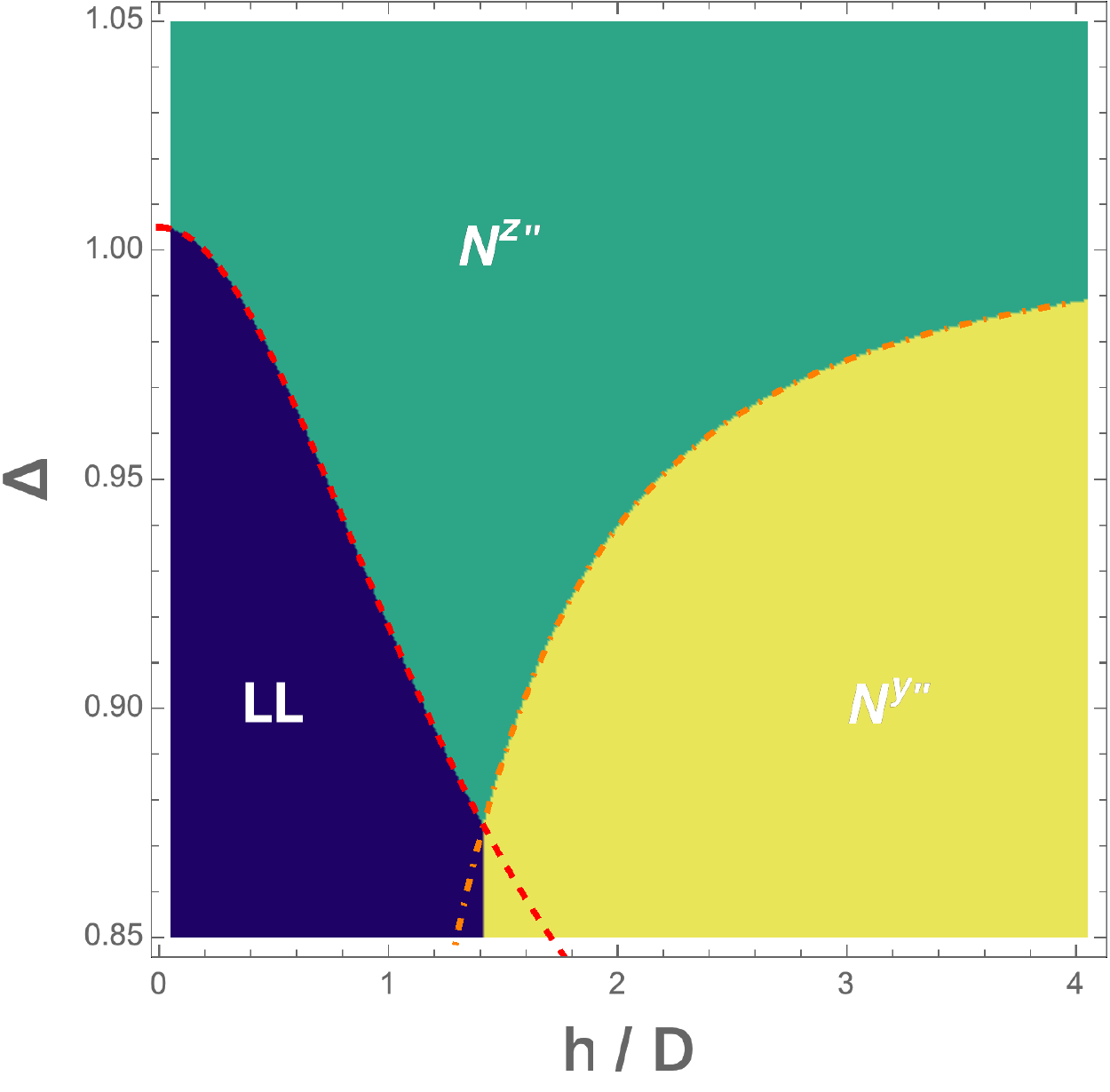}
	\caption{(Color online) Phase diagram for the case of relatively strong DM interaction $D/J=0.1$. Larger $D$ promotes $N^z$ state. The two phase boundaries are given
	by Eq.~\eqref{eq:deltac} (orange dot-dashed line), and Eq.~\eqref{eq:deltac2} (red dashed line). 
	The phase boundary between LL and $N^y$ is located at $h/D=\sqrt{2}$ and is independent of $\Delta$. }
	\label{fig:phase01}
\end{figure}
\begin{figure}[htb!]
	\centering
	\includegraphics[width=0.35\textwidth]{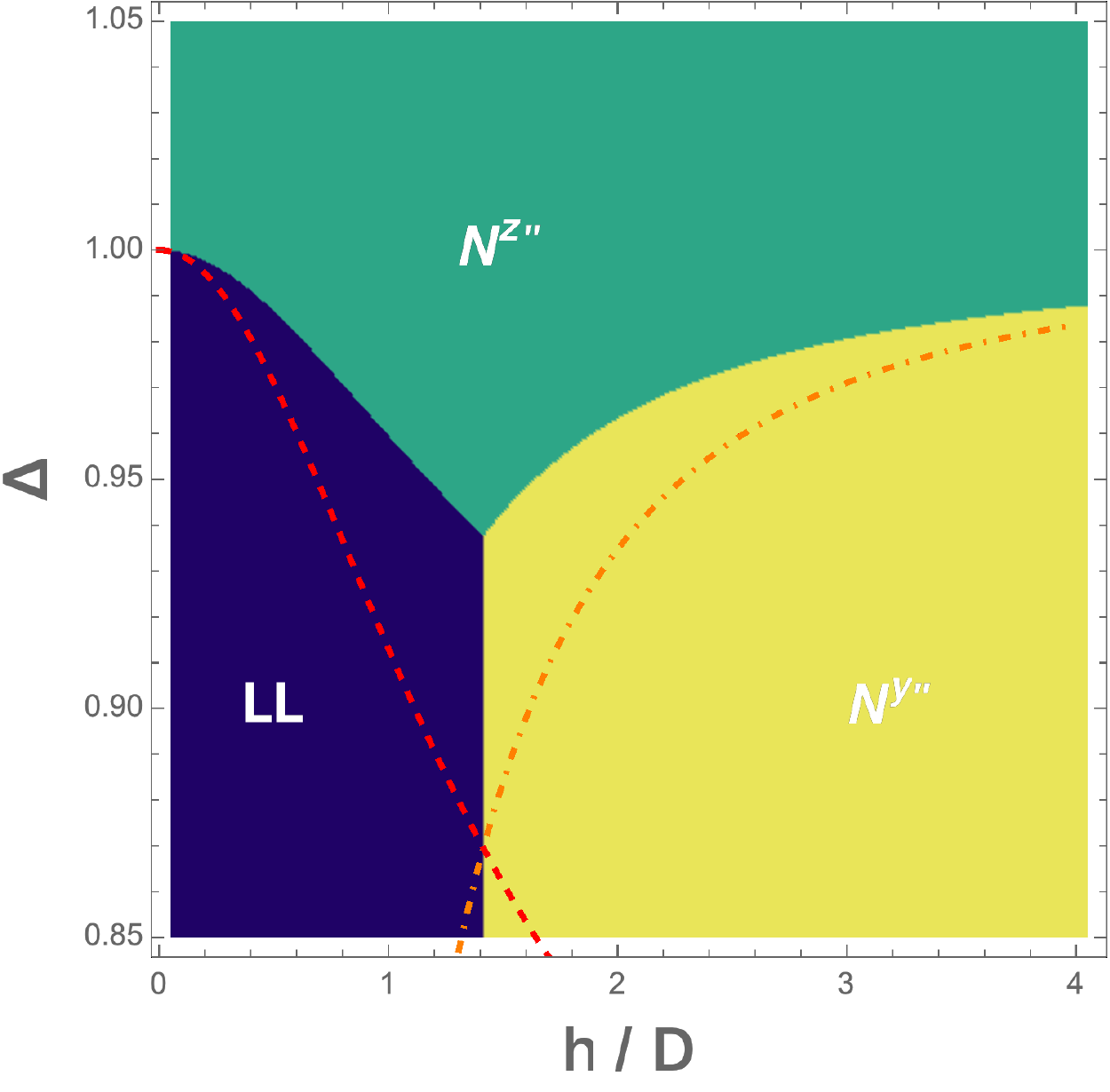}
	\caption{(Color online) Phase diagram for the case of a small DM interaction $D/J=0.01$ obtained via a two-stage RG process. At the first stage RG equations \eqref{eq:RG_bs} are integrated numerically. 
	Second-stage equations \eqref{eq:RG_tot2} are then solved analytically.
	The phase boundaries given by Eq.~\eqref{eq:deltac} 
	(orange dot-dashed line) and Eq.~\eqref{eq:deltac2} (red dashed line) are seen to deviate significantly from the actual ones.
	This shows the importance of the first-stage flow in the case of small and intermediate $D/J$.}
	\label{fig:phase001}
\end{figure}

The $\Delta-(h/D)$ phase diagrams are obtained by solving the RG equations and are presented in Figs.~\ref{fig:phase01} and \ref{fig:phase001}.
Figure~\ref{fig:phase01} is obtained under the condition that the first-stage RG flow can be skipped, due to the fact that $\ell_\varphi<0$ in Eq.~\eqref{eq:lphi}, which happens for sufficiently large $D$ and/or $h$.
Here we choose $D/J = 0.1$. In this situation we can determine the ground state simply by studying the initial conditions of the KT equations
according to the chart in Table~\ref{table:c} and Fig.~\ref{fig:kt}.

When  $\ell_\varphi>0$  oscillations develop over some finite lengthscale and one needs to integrate the first-stage RG equations~\eqref{eq:RG_bs} numerically for the interval $0 \leq \ell \leq \ell_\varphi$.
At the end of the first stage we obtain $y_C(\ell_\varphi)$, $y_\sigma(\ell_\varphi) $, and $C=y_\sigma^2(\ell_\varphi)-y_C^2(\ell_\varphi)$, which serve as initial values of the couplings 
for the second-stage, KT part, of the RG flow. This is the case of the $D/J=0.01$ phase diagram for which is presented in Fig.~\ref{fig:phase001}.

By comparing the phase diagrams in Figs.~\ref{fig:phase01} and \ref{fig:phase001}, we observe that large $D$ promotes the $N^z$ state, which is 
consistent with the numerical DMRG result in  Fig.~\ref{fig:phase}.

Next we study phase boundaries between different phases. 
\label{subsec:yandz}
Figure~\ref{fig:kt} shows that the phase transition between $N^y$ and $N^z$ states is related to the initial values of $y_C$ and $y_\sigma$.
The coupling $y_C(0)$ has opposite signs in the regions $3$ and $4$.  
Therefore in the $\Delta-h/D$ phase diagram this boundary  corresponds a critical value $\Delta_{c1}$ at which 
 $y_C(0)= 0$ and $C=y_\sigma^2(0)>0$.
These conditions indicate that the boundary is described by $D/h=\sqrt{\lambda/2}$, which leads to the explicit expression for it:
\begin{equation}
\Delta_{c1}=1+\frac{1}{2}(\frac{D}{J})^2-\frac{2}{c}(\frac{D}{h})^2.
\label{eq:deltac}
\end{equation}
For a fixed $D$, a larger field $h$ leads to a greater $\Delta_{c1}$, which is illustrated as an orange dot-dashed line in Figs.~\ref{fig:phase01} and \ref{fig:phase001}.
Figure \ref{fig:phase01} shows excellent agreement of the obtained phase transition line with the numerical solution of RG equations, due to the fact
that in this case the first stage of RG flow is not required. Interestingly, the limit of $D\to 0$, corresponding to $h/D \to \infty$ in the above figures,
is described by our theory as well, as we explain in Appendix~\ref{app:d=0}. In that case one deals with the $XXZ$ model in the transverse magnetic field for which
the critical line separating the two Ising phases $N^y$ and $N^z$ is reduced to the horizontal asymptote $\Delta_{c1}=1$, in agreement with the
previous study in Ref.~\onlinecite{Dmitriev2002}.

The boundary between the gapless LL and Ising $N^z$, according to Table \ref{table:c}, happens at $C=0$, $y_C(0)>0$, and $y_\sigma(0)<0$.
Therefore, we have the relation that $y_\sigma(0)=-y_C(0)$. 
This gives the critical $\Delta_{c2}$
\begin{equation}
\Delta_{c2}=1+\frac{1}{2}(\frac{D}{J})^2-\frac{2}{c}\frac{1}{1+2 (D/h)^2}.
\label{eq:deltac2}
\end{equation}
Therefore, in contrast to Eq.~\eqref{eq:deltac}, a larger field $h$ results in a smaller $\Delta_{c2}$. This result is also confirmed in Figs.~\ref{fig:phase01} and \ref{fig:phase001}. 

Finally, the transition between the LL and Ising $N^y$ is described by $C=0$, $y_C(0)<0$, and $y_\sigma(0)<0$. 
This gives $y_C(0)=y_\sigma(0)$, which is satisfied by $\cos\theta^-=1/3$ and $\lambda\geq1$. This condition implies that transition between the LL and $N^y$ is a 
vertical line located at $(h/D)_{c3}=\sqrt{2}$, which is again confirmed by numerical solution of the RG equations in Figs.~\ref{fig:phase01} and \ref{fig:phase001}. 
Different from the other two boundaries, the one between the LL and $N^y$ is independent of $\Delta$, and this is consistent with the classical analysis 
in Ref.~\onlinecite{Garate2010}. The constraint $\lambda\geq 1$ implies that this boundary exists only for $\Delta\leq \Delta_t\equiv1+(D/J)^2/2-1/c$. The crossing point 
of the critical lines $\Delta_{c1}$ and $\Delta_{c2}$ also gives the condition $(h/D)_{c3}=\sqrt{2}$. 
The triple point where three phases intersect is at $h/D=\sqrt{2}$ and $\Delta_t$. For $D/J=0.1$ in Fig.~\ref{fig:phase01} it is evaluated to be 
at $\Delta_t\simeq 0.874$. 

The main message of this section is that a strong DM interaction, acting jointly with the transverse magnetic field, causes significant modification of the classical phase diagram and works to stabilize Ising $N^z$ order 
well beyond its classical domain of stability (given by $\Delta > 1$), in agreement with the field-theoretic predictions of Refs.~\onlinecite{Gangadharaiah2008} and \onlinecite{Garate2010}.

\section{Numerical studies} 
\label{Sect:Numerics}

In this section, we will determine the ground-state properties of the model system in Eq.~(\ref{eq:H_0}) by an extensive and accurate DMRG\cite{White1992,White1993,Schollwock2005} simulation. Here we consider
a system with a total number of sites $L$ up to $L=1600$ and perform ten sweeps by keeping up to $m=400$ DMRG states with a typical 
truncation error of order $10^{-9}$.
In addition, we have also carried out an independent infinite time-evolving block decimation (iTEBD) \cite{Vidal2003,Vidal2004,Vidal2007} simulations with the same bond dimension and the same lengths as for the correlation function calculations. Our iTEBD results agree fully with our
DMRG results (see Fig.\ref{fig:phase} below).

Our principal results are summarized in the phase diagram in Fig.\ref{fig:phase} at $D/J=0.05$ and $D/J=0.1$. Changing the parameters $\Delta$ and $h/D$, we find three distinct phases, including a gapless LL phase and two ordered phases: the N\'eel Ising ordered $N^z$ (Ising order along the $z$ axis) and $N^y$ (Ising order along the $y$ axis) phases. Our numerical results show that the DM interaction stabilizes the $N^z$ Ising order which extends into the $\Delta<1$ region, while the $N^y$ Ising order gets suppressed by the DM interaction and gives way to the LL phase for a relatively small transverse magnetic field $h \lesssim D$. These results agree well with the field-theoretic predictions described in Secs.~\ref{sec:theory} and \ref{sec:theory-phase}, although with slightly different phase boundaries due to significant finite-size effects, which are 
described in more detail in Sec.~\ref{Sect:FiniteSizeEffect}.

\begin{figure}[t!]
\begin{center}
\includegraphics[width=8cm]{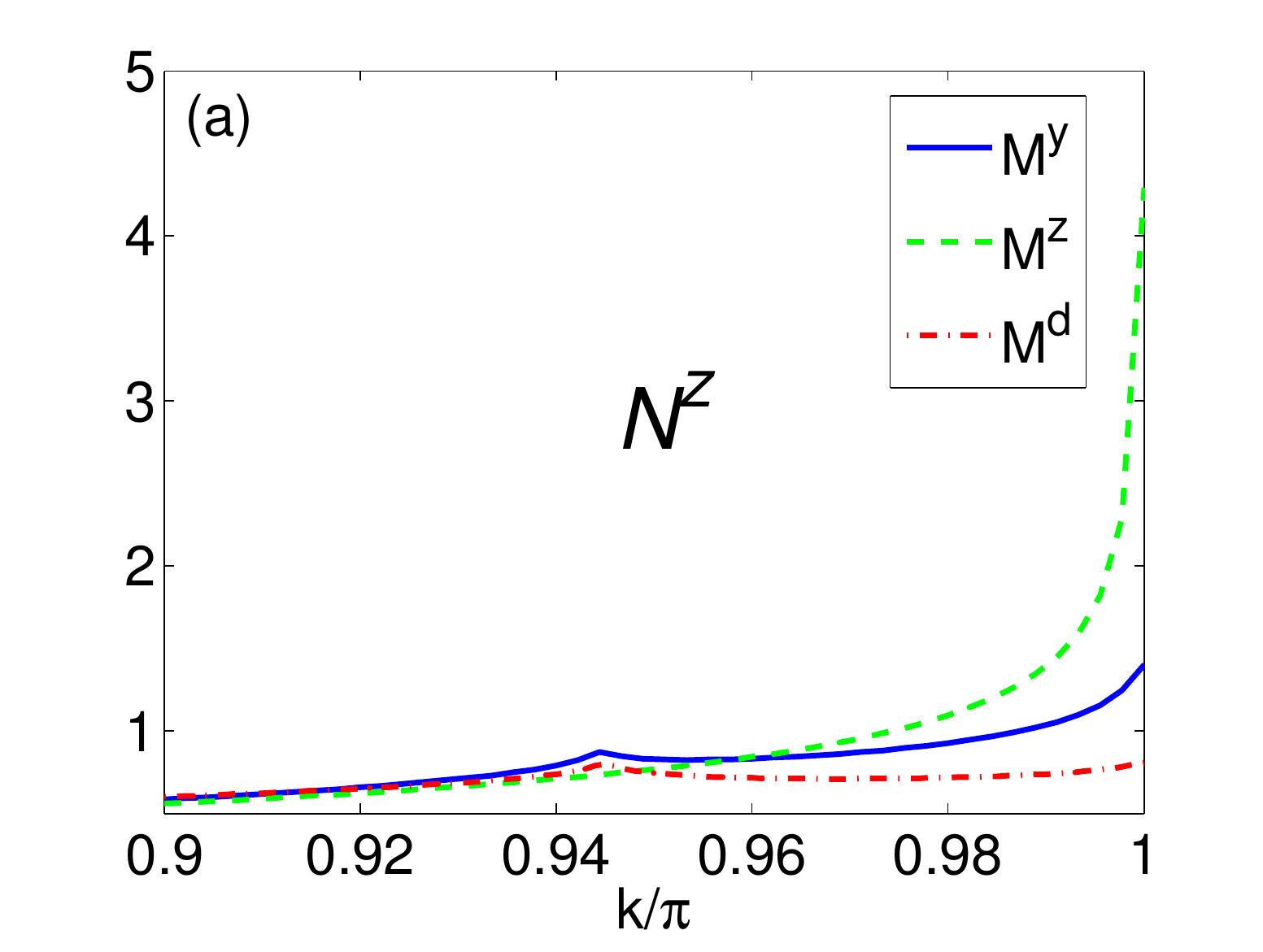}
\includegraphics[width=8cm]{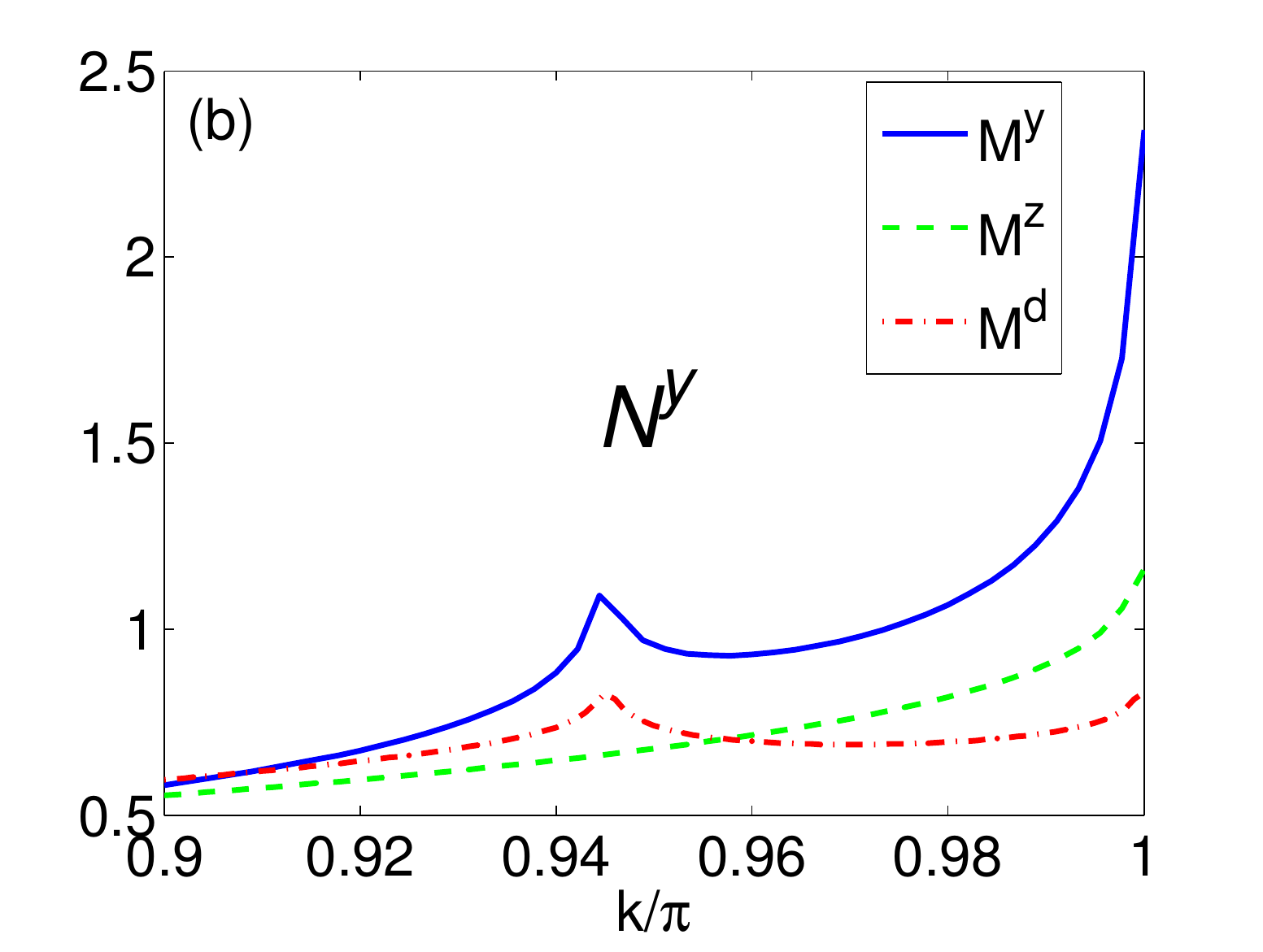}
\includegraphics[width=8cm]{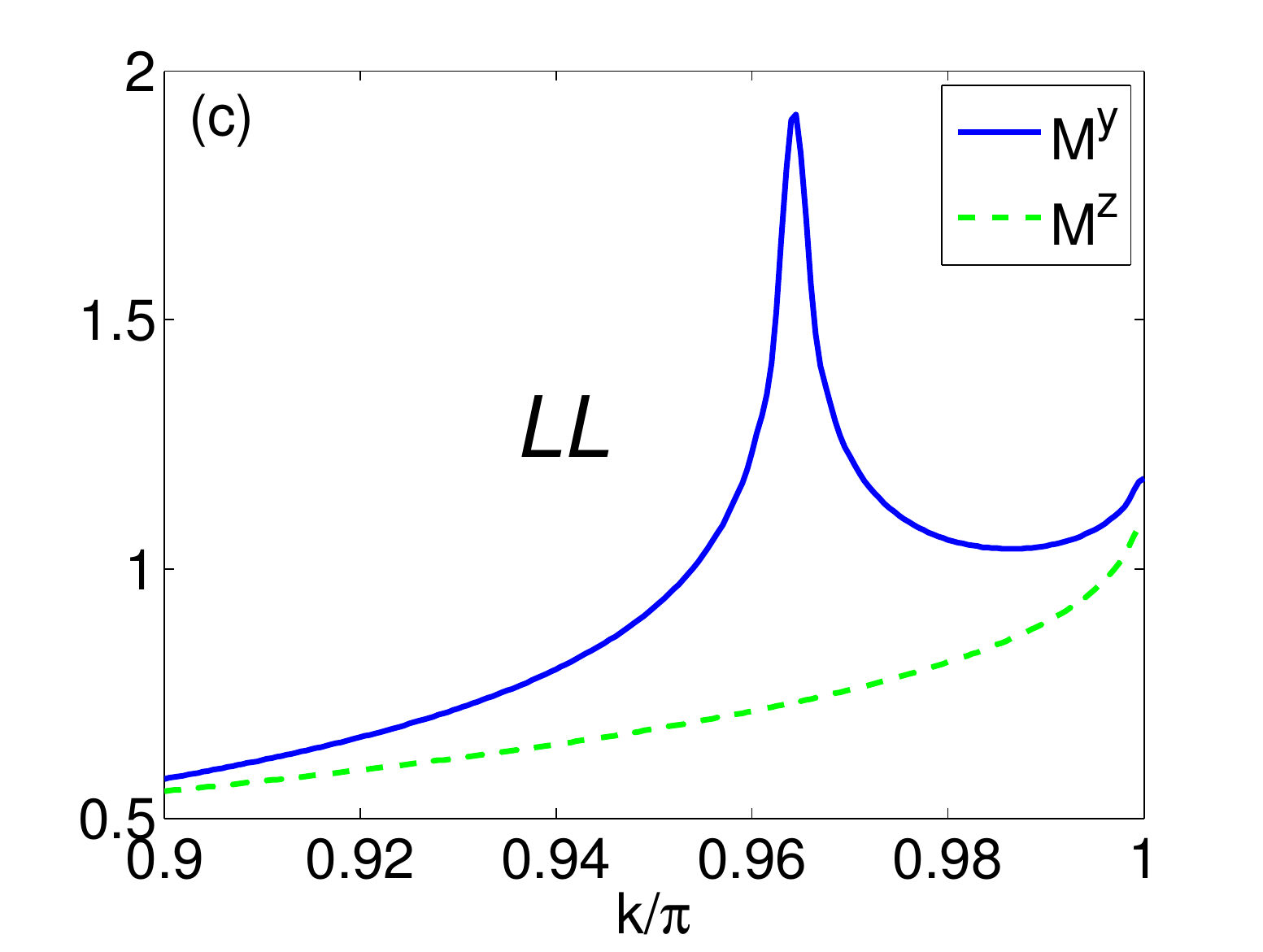}
\caption{(Color online) Structure factors $M_s^y$ (blue solid line) and $M_s^z$ (green dashed line) with $L=1600$ and $M_d$ (red dot-dashed line) with $L=1200$ (a) in the $N^z$ phase at $\Delta=1.1$, $D/J=0.1$, and $h/J=0.2$, (b) in the $N^y$ phase at $\Delta=0.7$, $D/J=0.1$, and $h/J=0.2$, and (c) in the $LL$ phase at $\Delta=0.7$, $D/J=0.1$, and $h/J=0.075$. 
} 
\label{fig:SF}
\end{center}
\end{figure}

To characterize distinct phases of the phase diagram, we measure magnetic correlations in the ground state by evaluating the equal time 
spin structure factor $M_s^\alpha(k)=\frac{1}{L}\sum^L_{ij}e^{\imath k (r_i-r_j)}\langle S^\alpha_iS^\alpha_j\rangle$, where $\alpha=x, y, z$ denotes different spin components. The structure factor is peaked at $k=\pi$ in both the $N^z$ and $N^y$ phases, corresponding to the commensurate N\'eel Ising order along the $z$ axis and $y$ axis, respectively. To quantitatively analyze this order, we perform an extrapolation of the spin order parameter $N^\alpha(k)=\sqrt{M^\alpha_s(k)/L}$ to the thermodynamic limit ($L=\infty$) according to the generally accepted form 
\begin{equation}
\label{eq:fit}
N^\alpha(k,L)=N^\alpha(k,\infty)+\frac{a}{\sqrt{L_{1/2}}}+\frac{b}{L_{1/2}}
\end{equation}
where $a$ and $b$ are fitting parameters (see Appendix~\ref{app:DMRG} for details). 
The structure factor for a finite system of length $L$ is calculated by using only the central $L_{1/2}=L/2$ part of finite systems. 
In addition to the spin order, we also calculate the dimer structure factor $M_d(k)=\frac{1}{L}\sum^L_{ij}e^{\imath k (r_i-r_j)}\langle B_i B_j\rangle$, where $B_i=\textbf{S}_i\cdot \textbf{S}_{i+1}$ denotes the bond operator (see Fig.~\ref{fig:SF} for an example of the
$M_d(k)$ data). Staggered dimerization $\epsilon(x)$, introduced in \eqref{eq:dimer}, represents the low-energy limit of the staggered part of the bond operator, $B_i \to {\cal B}(x) + (-1)^x \epsilon(x)$, while its uniform part ${\cal B}$ represents an average bond energy.

%%%%%%%%%%%DMRG phase diagram%%%%%%%%%%%%
\begin{figure}[t!]
\begin{center}
\includegraphics[width=9cm]{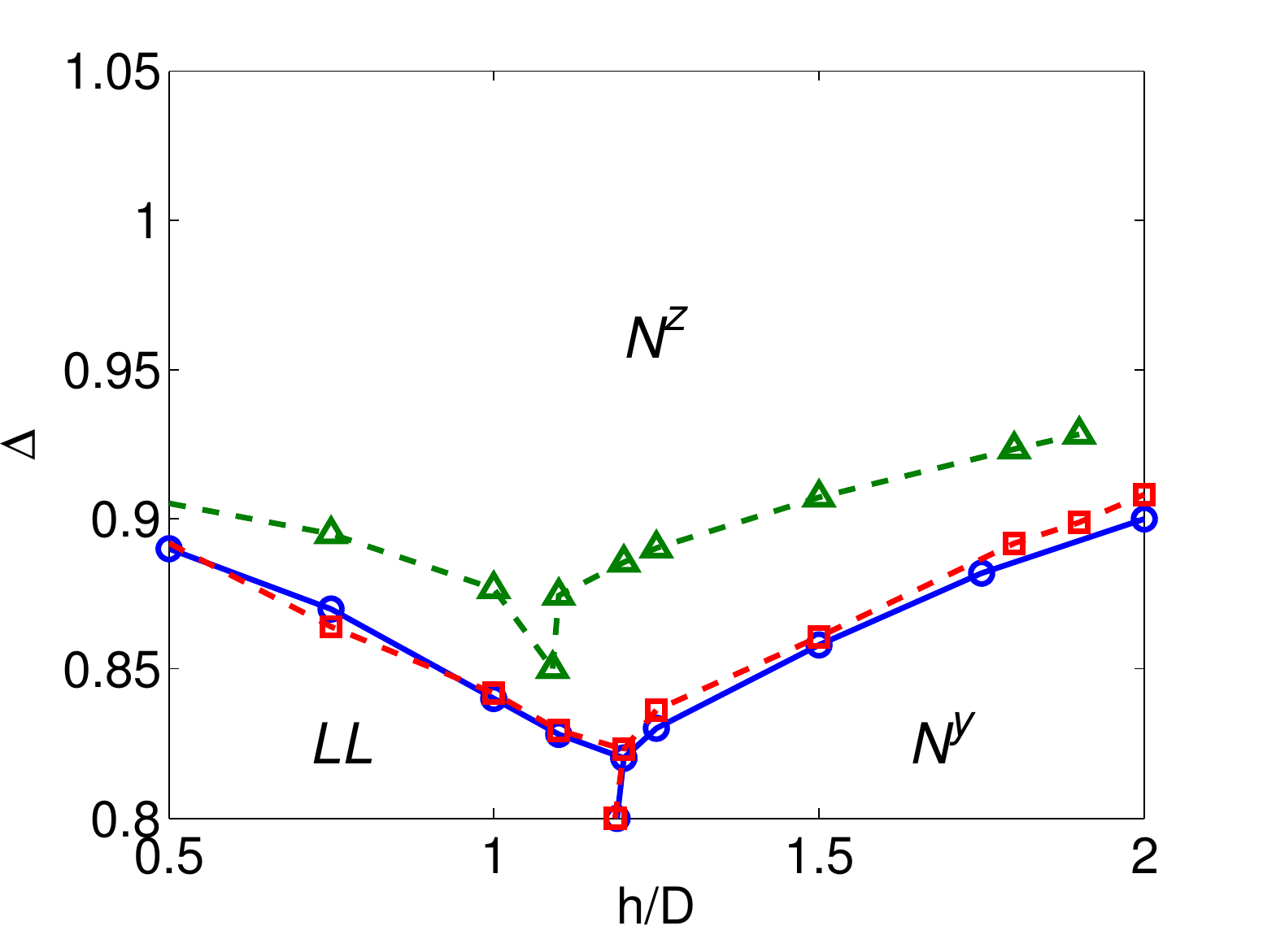}
\caption{(Color online) Ground-state phase diagram of the system in Eq.(\ref{eq:H_0}) at $D/J=0.1$ as determined by DMRG (open circles) and iTEBD (open squares) calculations for system with $L$=1200 sites. For comparison, the phase diagram at $D/J=0.05$ is also provided obtained by DMRG simulation for the system with $L=800$ sites (open triangles). The lines are guides for the eye.
} \label{fig:phase}
\end{center}
\end{figure}

To charaterize distinct phases in the phase diagram, we first show examples of both spin and dimer structure factors of the systems with length $L=1600$ in Fig.~\ref{fig:SF}.

\subsection{The $N^z$ phase}%%%%%%%%%
The $N^z$ phase is well understood for the case $\Delta>1$ without a DM interaction.
A finite DM interaction pushes the phase boundary to a lower $\Delta$ value due the renormalization of the effective anisotropy, which can be seen in the phase diagram. Figure~\ref{fig:SF}(a) plots the structure factors at $\Delta=1.1$, $D/J=0.1$, and $h/J=0.2$, where the structure factor $M^z_s(k)$ shows a clear peak at commensurate momentum $k=\pi$, indicating the presence of the N\`{e}el Ising order.
In contrast, the structure factor $M^y_s(k)$ has two smaller peaks, one at commensurate momentum $k=\pi$ and another at incommensurate momentum $k=k^\star<\pi$. 
However, since both peaks in the $M^y_s(k)$ structure factor are substantially smaller than the peak in $M^z_s(k)$ at commensurate $k=\pi$, 
we conclude that at this point the spin chain is the $N^z$ phase, with no $N^y$ kind of Ising order.

\subsection{The $N^y$ phase} %%%%%%%%%
When $\Delta$ is small while $h$ is sufficient large, the system enters into the $N^y$ phase. This phase is characterized by a dominant peak of the structure factor $M^y_s(k)$ at commensurate momentum $k=\pi$, while peaks in $M^z_s(k=\pi)$ and $M^y_s(k=k^\star)$ are much smaller [see Fig.~\ref{fig:SF}(b)]. Note that the $N^y$ N\'eel Ising order, which is also present in the system without a DM interaction, is suppressed by the finite DM interaction, especially for $h \leq D$. See Appendix~\ref{app:d=0} for an analytical explanation of this.

\subsection{The LL phase}%%%%%%%%%%%
The system is in the LL phase when both $\Delta$ and $h$ are small enough, and is characterized by the dominant peak in the structure factor $M^y_s(k)$ at the incommensurate momentum $k=k^\star<\pi$ as shown in Fig.~\ref{fig:SF}(c).  For example, the peak is at $k^*\approx 0.965\pi$ for $\Delta=0.7$, $D/J=0.1$, and $h/J=0.075$. For the same set of parameters, the field theory predicts the peak to be at $k=\pi \pm t_\varphi$, with $t_\varphi=\sqrt{h^2+D^2}/(\pi J/2)$ [see \eqref{shift:J-N} and \eqref{eq:LL4}]. This prediction translates into $k^*=0.975\pi$, which is consistent with the numerical result. 
Notice that our numerical calculations give a slightly smaller $k^*$, which is caused by the difference in spinon velocity $v$ 
from the zero field value $\pi J/2$ and finite-size effects.
Similar to $M^y_s(k)$, the dimer structure factor also exhibits a two-peak feature at both commensurate $k=\pi$ and incommensurate momenta $k=k^\star<\pi$. 
This is a direct consequence of the chiral rotation \eqref{chiral rotation}  which mixes up staggered magnetization and dimerization operators as Eqs. \eqref{eq:dimer} and \eqref{eq:Nrotation} 
[equivalently, \eqref{eq:Neps}] show.

Having characterized the distinct phases, now we can try to determine the phase boundary between them.
\subsection{The $N^y$-$N^z$ boundary}%%%%%%%%%%%
The phase boundary between the two Ising phases is determined by the order parameters $N^y(\pi)$ and $N^z(\pi)$, which should saturate to a finite nonzero value in the thermodynamic limit in the 
$N^y$ and $N^z$ phases, correspondingly, and vanish elsewhere.
 Unfortunately, due to large finite-size effects (see Sec. \ref{Sect:FiniteSizeEffect} for details), the order parameters tend to behave continuously across 
the anticipated phase boundary, even though their values in the ``wrong" phase become very small.
We therefore try to identify the phase boundary by looking for the crossing point where the two order parameters take the same value since the $N^z$ Ising order dominates at larger $\Delta$ 
while the $N^y$ order wins at smaller $\Delta$. An example of determining the phase boundary in this way is shown in Fig.~\ref{fig:order}(a) in the Appendix~\ref{app:DMRG}.

\subsection{LL-Ising boundary}%%%%%%%%%%%
In the LL phase all order parameters vanish in the thermodynamic limit. Unfortunately, again due to strong finite-size effects, an unambiguous identification of this phase is difficult 
since both Ising order parameters remain nonzero, although really small, inside it. We observe that in both $N^y$ and LL phases, the spin structure 
factor $M_s^y(k)$ develops peaks at commensurate momentum $k=\pi$ and at incommensurate momentum $k=k^\star<\pi$ (see Fig.~\ref{fig:SF}).
This is a direct consequence of Eqs.\eqref{eq:Neps} and \eqref{eq:Nrotation}, which show that 
$N^y \sim \cos\theta_0 {\cal N}^{y} + \sin\theta_0 \xi$.
While ${\cal N}^{y}$ is peaked at {\em zero} momentum [which means that its contribution to spin density 
$S^y \sim (-1)^x N^y$ is peaked at momentum $\pi$], the rotated dimerization operator $\xi$ is peaked at $\pm t_\varphi$ [see 
Eqs.\eqref{eq:shift} and \eqref{shift:J-N}]. Therefore, $M_s^y(k)$ is expected to have peaks at both $k=\pi$ {\em and} $k^* = \pi - t_\varphi$.
A similar two-peak structure, with maxima at momenta $\pi$ (coming from ${\cal N}^{y}$) and $k^*$ (coming from $\xi$), 
shows up in the dimer structure factor $M_d(k)$, in full agreement with the second line of \eqref{eq:Neps}. Figures~\ref{fig:SF} (a) and (b) show the
corresponding numerical data. 

Inside the $N^y$ phase the dominant peak of $M_s^y$ is at $k=\pi$, suggesting the well developed N\'eel order of the $N^y$ kind. 
In contrast, deep inside the LL phase, $M_s^y(k^*)$, which comes from power-law correlations of the rotated dimerization operator $\xi$, 
 dominates over the peak at $\pi$. This numerical finding is fully consistent with our low-energy bosonization calculation in Eq.~\eqref{eq:LL4},
 which shows that spin correlations caused by rotated operators $\xi$ and ${\cal N}^z$ are the slowest-decaying ones.
Therefore, the phase boundary between the LL and $N^y$ phases can be identified from the condition $M^y_s(k=k^\star) = M^y_s(k=\pi)$. 
The resulting phase boundary agrees well with the theoretical prediction. Similarly, the boundary between the LL and $N^z$ phases 
is determined by 
$M^y_s(k=k^\star) = M^z_s(k=\pi)$, [see Fig.~\ref{fig:order}b].
Since $M^y_s$ shows a dominant peak at $k=k^\star$ in the LL phase while the ${\cal N}^z$ phase has a dominant order at $k=\pi$, the phase boundary between these two phases can be determined by the crossing point of the above quantities.

\begin{figure}[t]
\begin{center}
\includegraphics[width=9cm]{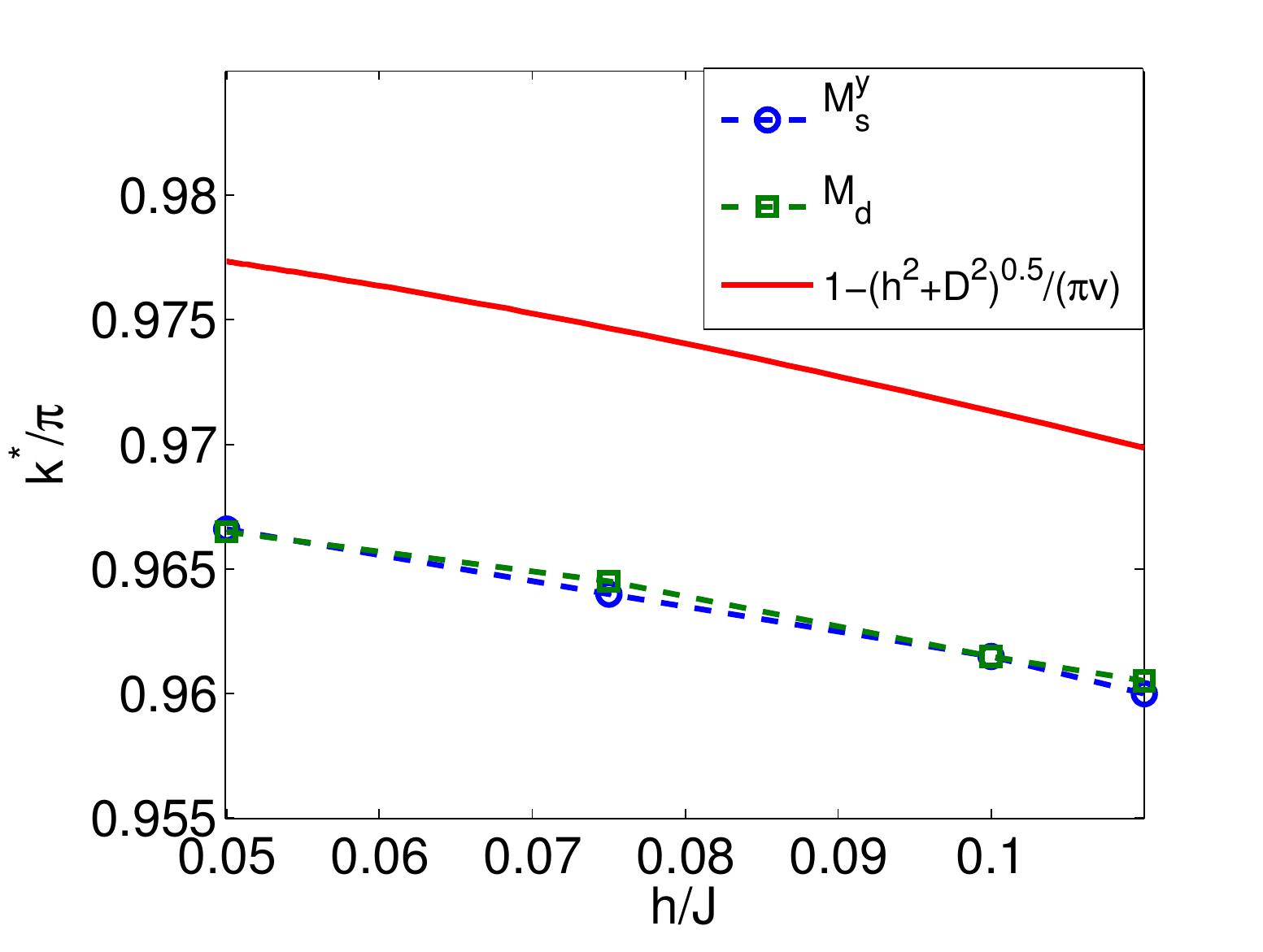}
\caption{(Color online) Dependence of the incommensurate peak momentum $k^\star$ in the spin structure factor $M^y_s(k)$ (blue circles) and dimer structure factor $M_d(k)$ (green squares) as a function of the transverse magnetic field $h$ at $D/J=0.1$. The red line denotes the theoretical prediction, we used $v=\pi J/2$. }
\label{fig:kstar}
\end{center}
\end{figure}

Further quantitative agreement can be established by comparing numerical data for $k^\star$, extracted from $M_s^y(k)$ and $M_d(k)$ data,
with the analytical prediction $k^* = \pi - t_\varphi = \pi - \sqrt{D^2 + h^2}/v$, as shown in Fig.~\ref{fig:kstar}. The small difference between the measured
and the predicted $k^*$ values is probably due to our omission of the velocity renormalization by marginal operators.

Finally, we have also calculated the phase diagram of the system with a smaller DM interaction $D/J=0.05$. The phase diagram for the $L=800$ chain 
is shown in Fig.~\ref{fig:phase} by a green dashed line. Compared with the larger DM interaction $D/J=0.1$ case, the phase boundaries for both the
$N^z$-$LL$ and $N^z$-$N^y$ phase transitions move to higher $\Delta$ values, in qualitative agreement with theoretical expectations (see
phase diagrams in Figs.~\ref{fig:phase01} and \ref{fig:phase001} for a similar comparison).

\section{Analytical understanding of finite size effects in DMRG study} \label{Sect:FiniteSizeEffect}

Our formulation provides a convenient way to understand some of the finite-size effects unavoidable in the numerical study of the problem.
Here we focus on the case of a relatively strong DM interaction $D/J=0.1$, analytical and numerical phase diagrams for which are presented in Figs.~\ref{fig:phase01} and~\ref{fig:phase}, correspondingly.

\begin{figure}[!]
	\centering
	\includegraphics[width=0.4\textwidth]{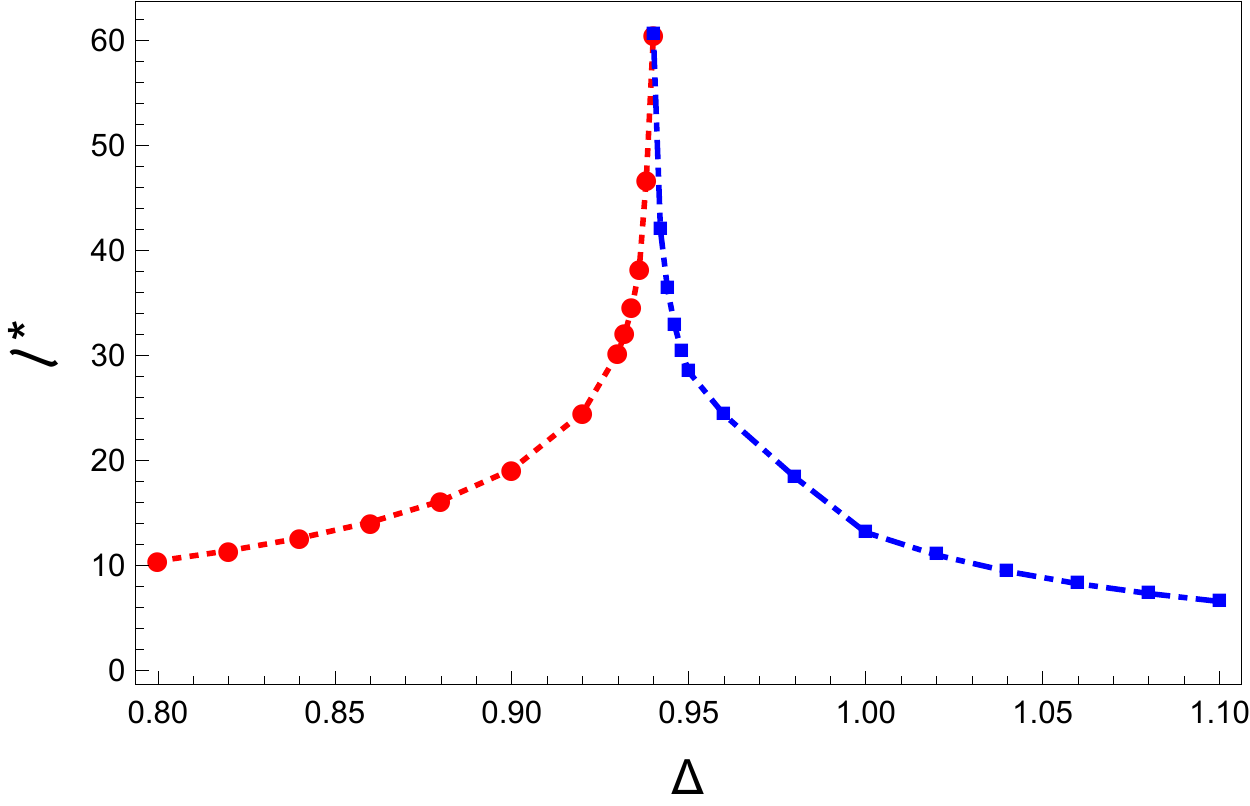}
	\caption{(Color online) Critical RG $\ell^*$ for which $|y_c(\ell^*)|=1$ (obtained by solving the KT equations; see Appendix~\ref{app:ktsolution}) as a function of the $XXZ$ anisotropy $\Delta$. Here $D/J=0.1$ and $h/J=0.2$. The system is in the $N^y$ phase (red line) for $\Delta<\Delta_c\simeq0.94$ [the phase boundary $\Delta_c$ is determined from Eq.~\eqref{eq:deltac}], while at $\Delta>\Delta_c$ the system enters the $N^z$ phase (blue line). 
	Near the transition point, $\ell^*\gg \ell_s =7.37$.}
	\label{fig:lstar}
\end{figure}

By solving the RG equations \eqref{eq:RG_tot2} we obtain the critical RG scale $\ell^*$ at which the order develops fully, namely, $|y_{C}(\ell^*)|=1$.
We find that $\ell^*$ grows rapidly as $\Delta$ approaches the phase boundary between the $N^y$ and $N^z$ states, as shown in Fig.~\ref{fig:lstar}, with
$\ell \approx 50$ near the critical point.
However the finite size of the system used in the DMRG study, $L= 1600$ in units of the lattice spacing $a$, corresponds to a much smaller RG scale 
of $\ell_s = \ln[1600] = 7.37$.
Therefore the RG scales greater than $\ell_s$ are {\em not accessible} for the DMRG. In other words, if we associate the correlation length $\xi = a e^{\ell^*}$
with the order which develops at $\ell^*$, and if it happens that $\ell^*> \ell_s = 7.37$, then the DMRG simulations will not be sensitive
to the development of the long-range order in this case. This is the basic explanation of the unavoidable difficulty one encounters 
in numerical determination of the phase boundaries between various phases.

\begin{figure}[!]
	\centering
	\includegraphics[width=0.4\textwidth]{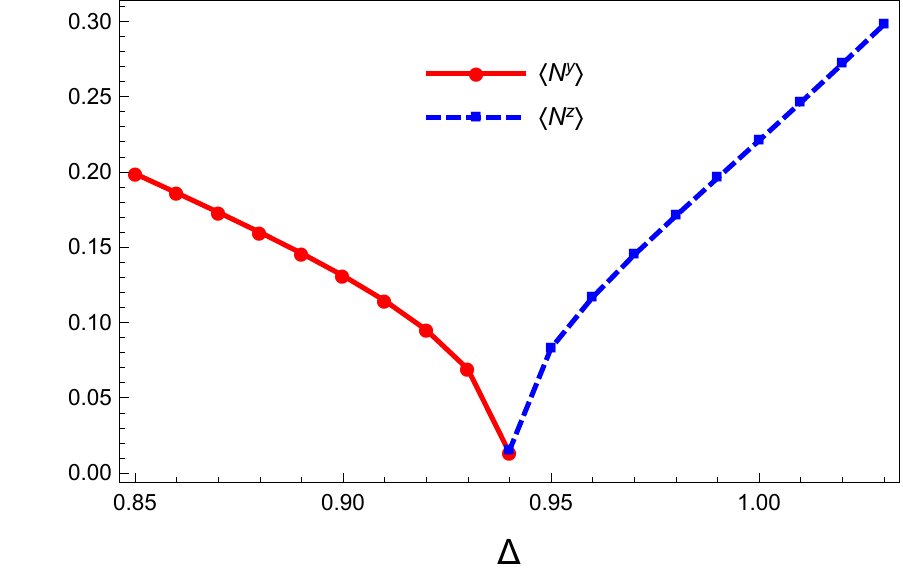}
	\caption{(Color online) Order parameters as a function of $\Delta$ for two ordered states $N^y$ and $N^z$, at $D/J=0.1$, $h/J=0.2$, and 
	RG length scale $\ell =\ell_s$. Here $\Delta$ is near the phase boundary $\Delta_c\simeq0.94$ (determined from Eq.~\eqref{eq:deltac}). See the main text and Appendix \ref{app:orderp} for details.}
	\label{fig:orderp}
\end{figure}

In addition to calculating the $\ell^*$ associated with the development of long-range order, we can also calculate the order parameters for the $N^y$ and $N^z$
phases developing in the system as functions of the running RG scale $\ell$. Appendix~\ref{app:orderp} describes how it is done. 
We show there that the required order parameters are given by 
 \begin{equation}
 \begin{gathered}
 \langle N^y\rangle= \langle \text{Re}[e^{i\beta\vartheta/2}]\rangle, \quad\langle N^z\rangle= \langle \text{Im}[e^{i\beta\vartheta/2}]\rangle.
 \end{gathered}
 \end{equation}
Equation~\eqref{eq:expec} shows the explicit form of the order parameters in terms of running couplings $y_{C,\sigma}(\ell)$.
Figure~\ref{fig:orderp} illustrates our results. It shows the order parameters $ \langle N^{y,z}\rangle$ which are evaluated at the maximum possible for our chain RG scale $\ell = \ell_s$.
Observe that, in agreement with the numerical data in Fig.~\ref{fig:order}(a), there is a noticeable asymmetry between these two order parameters:
The order parameter of the $N^y$ phase is smaller than that of the $N^z$ phase.

\section{Conclusions}

Our extensive DMRG study shows an excellent agreement with the analytical investigation based on the RG analysis of the weakly perturbed Heisenberg chain.
We have worked out a full phase diagram of the model in the $\Delta-(h/D)$ plane. Our numerical findings match predictions of Ref.~\onlinecite{Garate2010} well and
confirm the prevalence of $N^z$ N\'eel Ising order in the regime of comparable DM and magnetic field magnitudes \cite{Gangadharaiah2008}.
In addition, we find that significant finite-size corrections observed numerically are well explained by the logarithmic slowness of the KT RG flow.
As a result of that, very large RG scales $\ell^*$, far exceeding those set by the finite length $L$ of the chain used in the DMRG, are required to reach the Ising-ordered phases.

Our numerical data also confirm the existence of the critical Luttinger liquid phase with fully broken spin-rotational invariance. 
This phase with dominant incommensurate spin
and dimerization power-law correlations is a quantum analog of the classical chiral soliton lattice.

Our findings open up the possibility of an experimental check of theoretical predictions in quasi-one-dimensional antiferromagnets 
with a uniform DM interaction \cite{Halg2014,Smirnov2015}.
The idea is to probe the spin correlations at a finite temperature above the critical ordering temperature of the material when interchain spin correlations, 
which drive the three-dimensional ordering, are not important while individual chains still possess anisotropy of spin correlations, sufficient for experimental detection, caused by the uniform DM interaction.
Under these conditions one should be able to probe the fascinating competition between the uniform DM interaction and the transverse 
external magnetic field.

\begin{acknowledgments}
O.A.S. would like to thank I. Affleck for several discussions related to this work. W.J. and O.A.S. were supported by the National Science Foundation Grant No. NSF DMR-1507054. H.-C.J was supported by the Department of Energy, Office of Science, Basic Energy Sciences, Materials Sciences and Engineering Division, under Contract No. DE-AC02-76SF00515. Y.-H. C. was supported by a Thematic Project at Academia Sinica.
\end{acknowledgments}

%%%%%%%%%%%%%%%%%%%%Appendix%%%%%%%%%%%%%%%%%%%
\appendix

\renewcommand\thefigure{S\arabic{figure}}
\setcounter{figure}{0}

\section{Bosonization}\label{app:bosonization}
The low-energy description is provided by the parametrization\cite{Gangadharaiah2008} ${\bm S}(x) \approx {\bm J}(x) + (-1)^n {\bm N}(x)$, where ${\bm J} = {\bm J}_L + {\bm J}_R$, with
${\bm J}_{L}(x)$ and ${\bm J}_{R}(x)$ are the uniform left- and right-moving spin currents, and ${\bm N}(x)$ is the staggered magnetization 
(our order parameter). Here $x=n a$ in terms of lattice constant $a$. These fields are expressed in terms of bosonic fields $(\phi,\theta)$
[this expansion is not specific to the SU(2), Heisenberg, point and can be generalized easily to a more general $XXZ$ Hamiltonian],
\begin{equation}
\begin{gathered}
J_R^{+}=\frac{1}{2\pi a}e^{-i\sqrt{2\pi}(\phi-\theta)},\;
J_L^{+}=\frac{1}{2\pi a}e^{i\sqrt{2\pi}(\phi+\theta)},\\
J_R^{z}=\frac{\partial_x\phi-\partial_x\theta}{2\sqrt{2\pi}},\;
J_L^{z}=\frac{\partial_x\phi+\partial_x\theta}{2\sqrt{2\pi}},
\end{gathered}
\label{eq:J}
\end{equation}
and
\begin{equation}
{\bm N}=A(-\sin[\sqrt{2\pi}\theta ],\; \cos[\sqrt{2\pi} \theta],\;
-\sin[\sqrt{2\pi}\phi]).
\label{eq:staggered_m}
\end{equation}
Here, $A\equiv \gamma/\pi a_0$ and $\gamma=\langle \cos(\sqrt{2\pi}\phi_{\rho})\rangle \sim O(1)$ is 
determined by gapped charged modes of the chain.
The Hamiltonian in Eq.~\eqref{eq:H_0} is approximated in the low energy limit as \cite{Gangadharaiah2008,Schnyder2008,Garate2010}
\begin{equation}
H=H_0  + V  + H_{\rm bs},
\label{system_1}
\end{equation}
where
\begin{equation}
\begin{split}
H_0&=\frac{2\pi v}{3}\int \mathrm{d}x ({\bm {J}_{R} }\cdot{\bm {J}_{R} }+{\bm {J}_{L} }\cdot{\bm {J}_{L} }),\\
V&=-{D} \int \mathrm{d}x(J_{R}^{z}-J_{L}^{z})-h\int \mathrm{d}x(J_{R}^{x}+J_{L}^{x}),\\
H_{\rm bs}&=-g_{\rm bs}\int \mathrm{d}x [ J_R^xJ_L^x+J_R^yJ_L^y+(1+\lambda)J_R^zJ_L^z ],\\
\end{split}
\label{system_2}
\end{equation}
where $\lambda$ is the total $XXZ$ anisotropy described by Eq.~\eqref{eq:lambda}. 

\section{Chiral rotation}
\label{app:chiral rotation}
The system Hamiltonian is described in Eq.~\eqref{system_2}. It is convenient to exploit the extended symmetry of $H_0$ and treat both vector perturbations $h$ and $D$ equally by performing a chiral rotation of spin currents about the ${y}$ axis 
\cite{Gangadharaiah2008,Schnyder2008,Garate2010}
\begin{equation}
{\bm J}_{R/L}={\cal R}(\theta_{R/L}){\bm M}_{R/L},
\label{chiral rotation}
\end{equation}
with ${\bm M}_{R/L}$ is the spin current in the rotated frame, and $\cal R $ is the rotation matrix,
\begin{equation}
\begin{gathered}
{\cal R}(\theta_{R/L})=
\begin{pmatrix}
\cos\theta_{R/L} & 0 & \sin\theta_{R/L} \\
0 & 1 & 0 \\
-\sin\theta_{R/L} & 0 & \cos\theta_{R/L}.
\end{pmatrix},
\end{gathered}
\end{equation}
where
\begin{equation}
\begin{gathered}
\theta_R=\theta_0+\pi/2,\; \theta_L=-\theta_0+\pi/2,\;\theta_0\equiv\arctan\left(\frac{-D}{h}\right).
\end{gathered}
\end{equation}

%OS
\begin{comment}
{\bf OS} I think it is better to re-write these equations as 
\begin{equation}
\begin{gathered}
\theta_R=\pi/2 - \theta_0,\; \theta_L=\pi/2 + \theta_0,\;\theta_0\equiv\arctan\left(\frac{D}{h}\right).
\end{gathered}
\end{equation}
\end{comment}

Via this chiral rotation, 
 vector perturbation $V$ in Eq.~\eqref{system_2} becomes
\begin{equation}
\begin{split}
V&=-\sqrt{D^2+h^2}\int \mathrm{d}x (M_{R}^z+M_{L}^z)\\
&=-\frac{\sqrt{D^2+h^2}}{\sqrt{2\pi}}\int \mathrm{d}x\partial_x\varphi.
\end{split}
\label{eq:effV}
\end{equation}
The staggered magnetization transforms as 
\begin{equation}
{\bm N}= (-{\cal N}^{z}, \cos\theta_0 {\cal N}^{y} + \sin\theta_0 \xi, {\cal N}^{x}),
\label{eq:Nrotation}
\end{equation}
Here ${\bm {\mathcal N}}$ and $\xi$ denote the staggered magnetization and dimerization in the rotated frame. They, as well as rotated spin currents ${\bm M}_{R/L}$, are expressed in terms of Abelian bosonic fields $\varphi$ and $\vartheta$. Staggered magnetization ${\bf N}$ in \eqref{eq:staggered_m}, staggered dimerization $\epsilon=({\gamma}/{\pi a_0})\cos[\sqrt{2\pi}\phi]$, and
spin currents ${\bm J}_{R/L}$ are written in terms of a $(\phi, \theta)$ pair, as Eqs.~\eqref{eq:J} and \eqref{eq:staggered_m} show.
Therefore, in the rotated frame
 \begin{equation}
{\bm {\mathcal N}}= \frac{\gamma}{\pi a_0} (-\sin\sqrt{2\pi} \vartheta, \cos\sqrt{2\pi} \vartheta, -\sin\sqrt{2\pi} \varphi)
\label{eq:N2}
\end{equation}
and $\xi = ({\gamma}/{\pi a_0})\cos\sqrt{2\pi} \varphi$.

The relation \eqref{eq:Nrotation} is obtained by observing that chiral rotation \eqref{chiral rotation} of vector currents corresponds to 
 the following rotation of Dirac spinors \cite{Garate2010,starykhbook2010} $\Psi_{R/L, s} = e^{-i\theta_{R/L} \sigma^y/2} \tilde{\Psi}_{R/L, s}$ in terms of which 
 spin currents are expressed \cite{ope} as $J^a_{R/L} = \Psi_{R/L}^+ \sigma^a \Psi_{R/L}/2$ and $M^a_{R/L} = \tilde{\Psi}_{R/L}^+ \sigma^a \tilde{\Psi}_{R/L}/2$.
 The (original) staggered magnetization, $N^a = (\Psi_{R}^+ \sigma^a \Psi_{L} + \Psi_{L}^+ \sigma^a \Psi_{R})/2 $, rotates into \eqref{eq:Nrotation}.
 Similarly, staggered dimerization $\epsilon(x) \sim (-1)^{x/a} {\bf S}(x) \cdot {\bf S}(x+a)$ transforms as 
 \begin{equation}
 \epsilon = \cos\theta_0 \xi - \sin\theta_0 {\cal N}^{y}.
 \label{eq:dimer}
 \end{equation}

The rotation \eqref{chiral rotation} transforms the backscattering Hamiltonian in \eqref{system_2} into,
 \begin{equation}
 \begin{split}
{H}_{\rm bs}=&2\pi v\int \mathrm{d}x \Big[\sum_{\alpha}y_\alpha M_R^{\alpha}M_L^{\alpha}+\\
&+y_A(M_R^zM_L^x-M_R^xM_L^z)\Big],
 \end{split}
\label{eq:hbs}
 \end{equation}
 where $\alpha=x,y,z$ and the initial values of coupling constants $y_\alpha$ and $y_A$ are shown in Eq.~\eqref{initial}.
 \begin{comment}
 \begin{equation}
 \begin{gathered}
 y_x(0)=-\frac{g_{\rm bs}}{2\pi v}[(1+\frac{\lambda}{2})\cos\theta^- +\frac{\lambda}{2}],\\
 y_y(0)=-\frac{g_{\rm bs}}{2\pi v},\\
 y_z(0)=-\frac{g_{\rm bs}}{2\pi v}[(1+\frac{\lambda}{2})\cos\theta^- -\frac{\lambda}{2}],\\
 y_A(0)=\frac{g_{\rm bs}}{2\pi v}(1+\frac{\lambda}{2})\sin\theta^-,\qquad
 \end{gathered}
 \end{equation}
 where the angle parameter is
 \begin{equation}
 \begin{gathered}
 \theta^-=2\theta_0,\quad \theta_0=\arctan(-D/h).
 \end{gathered}
 \end{equation}
\end{comment}

 We see from Eq.~\eqref{eq:effV} that in the rotated frame the chain experiences 
 an external magnetic field $h_{\rm eff} \equiv \sqrt{D^2+h^2}$ applied along the $z$ axis. 
This term is then absorbed into the isotropic Hamiltonian $H_0$ by the position-dependent shift
\begin{equation}
\varphi \to \varphi + t_\varphi x, \quad
t_{\varphi}\equiv\sqrt{D^2+h^2}/v=h_{\rm eff}/v.
\label{eq:shift}
\end{equation} 
As a result of this shift, the spin currents, the staggered magnetization and the dimerization in the rotated frame are modified as 
\begin{equation}
\begin{gathered}
M_R^{+}\to M_R^{+}e^{-it_\varphi x},\quad
M_L^{+}\to M_L^{+}e^{it_\varphi x},\\
M_R^{z}\to M_R^{z}+\frac{t_\varphi}{4\pi},\quad
M_L^{z}\to M_L^{z}+\frac{t_\varphi}{4\pi},
\end{gathered}
\end{equation}
and
\begin{equation}
\begin{gathered}
{\cal N}^{z} \to -\frac{\gamma}{\pi a_0}\sin(\sqrt{2\pi}\varphi+t_\varphi x),\\
\xi\to\frac{\gamma}{\pi a_0}\cos[\sqrt{2\pi}\varphi+ t_\varphi x].\\
\end{gathered}
\label{shift:J-N}
\end{equation}
The $\varphi$ field shift~\eqref{eq:shift} will also  transform the expression for the chain backscattering \eqref{eq:hbs} to Eq.~\eqref{eq:Hbs}, 
in which we neglected additional small terms coming from the shifts in $M_{R/L}^z$. 

\section{Analytical solution of Kosterlitz-Thouless (KT) equations}
\label{app:ktsolution}
Analytical solution of the KT equations~\eqref{eq:RG_tot2} is given by
\begin{equation}
y_\sigma(l)=
\begin{dcases}
\mu\dfrac{y_\sigma(0)\cosh(\mu l)-\mu\sinh(\mu l)}{-y_\sigma(0)\sinh(\mu l)+\mu\cosh(\mu l)},&\; C>0,\\
\mu\dfrac{y_\sigma(0)\cos(\mu l)+\mu\sin(\mu l)}{-y_\sigma(0)\sin(\mu l)+\mu\cos(\mu l)}, &\; C<0.\\
\end{dcases}
\label{eq:gen-sol}
\end{equation}
with $\mu=\sqrt{|C|}$. Also,
\begin{equation}
y_C(l)={\rm sgn}[y_C(0)]\sqrt{y_\sigma(l)^2-C}.
\label{eq:ycl}
\end{equation}
The sign of $y_C(l)$ depends on the sign of its initial value. The critical $\ell^*$, at which $|y_C(l=l^*)|=1$, can be determined by Eqs.~\eqref{eq:gen-sol} and \eqref{eq:ycl}, and is shown is Fig.~\ref{fig:lstar}.

\section{The $XXZ$ model in transverse field, $D=0$}
\label{app:d=0}
If we set $D=0$, two rotation angles $\theta_R=\theta_L=\pi/2$, and $\theta^-=0$. Then $y_A(0)=0$. 
In this condition, our model Hamiltonian~\eqref{eq:H_0} reduces to a $XXZ$ model in a uniform transverse field.  
The RG equations for the backscattering are,
\begin{equation}
\begin{gathered}
\frac{dy_x}{dl}=y_yy_z,\;\;
\frac{dy_y}{dl}=y_x y_z,\;\;
\frac{dy_z}{dl}=y_x y_y,\\
\end{gathered}
\end{equation}
and the initial values are,
\begin{equation}
\begin{gathered}
y_x(0)=-\frac{g_{\rm bs}}{2\pi v}[1+\lambda],\;
y_y(0)=y_z(0)=-\frac{g_{\rm bs}}{2\pi v},
\end{gathered}
\end{equation}
It is easy to find that $y_y(\ell) = y_z(\ell)$ for all $\ell$, so the RG equations above again acquire a KT form.
Now $\lambda=c(1-\Delta)$, so we obtain
\begin{equation}
\begin{gathered}
y_C(0)=-\frac{g_{\rm bs} }{4\pi v}\lambda,\; y_\sigma(0)=\frac{g_{\rm bs}}{2\pi v},\;
C=(\frac{g_{\rm bs}}{2\pi v})^2(1-\frac{\lambda^2}{4}).
\end{gathered}
\end{equation}
Using Eq.~\eqref{eq:gen-sol}, we find
\begin{equation}
y_\sigma(\ell) = 2\mu \frac{y_C^2/(y_\sigma + \mu)^2}{e^{-2\mu\ell} - y_C^2/(y_\sigma + \mu)^2},
\end{equation}
where the $y_{C/\sigma}$ on the right-hand-side are those at $\ell=0$ (their initial values). Therefore, since 
$y_\sigma(0) = {g_{\rm bs}}/{(2\pi v)} > \mu = \sqrt{y_\sigma^2 - y_C^2}$,
there is a divergence, signaling a strong-coupling limit, at $\ell_{\rm div} \approx \mu^{-1} \ln\left[4 |\lambda|^{-1}\right]$.
Observe that $\ell_{\rm div}$ is finite for any $\Delta \neq 1$, meaning that the two ordered phases are separated by the critical LL one,
which is just an isotropic Heisenberg chain in a magnetic field.

For $\Delta<1$, we have $\lambda>0$, $y_C(0)<0$, and then $y_C(l)\to -\infty$,  which leads to the $N^y$ state. 
For $\Delta>1$, instead $\lambda<0$ and $y_C(0)>0$, so that $y_C(l)\to +\infty$, one obtains the $N^z$ state. 
These two phases are separated by the critical line at $\Delta=1$.
Our phase diagrams in Figs.~\ref{fig:phase001} and ~\ref{fig:phase01} display exactly this behavior: Setting $D=0$ places the model at $h/D\to \infty$, where the critical line
separating the two Ising states approaches a horizontal asymptote at $\Delta =1$.

The above argument agrees with Ref.~\cite{Dmitriev2002}, which studied the ground state of the Hamiltonian
\begin{equation}
{\cal H}=\sum_{j} \left[J( {S}^{x}_{j} {S}^{x}_{j+1}+{S}^{y}_{j} {S}^{y}_{j+1}+\Delta {S}^{z}_{j}{S}_{j+1}^{z}) - hS^x_j\right].
\label{eq:xxz}
\end{equation}
It was found that for $h\neq 0$ the spectrum is gapped for both $\Delta >1$ and $\Delta < 1$ \cite{Dmitriev2002}.
The Ising order that develops is of the $N^z$ ($N^y$) kind for $\Delta >1$ ($\Delta < 1$).
Our RG equations evidently capture this physics well.

\section{Calculation of the order parameter}
\label{app:orderp}

In Ref.~\onlinecite{Lukyanov1997} Lukyanov and Zamolodchikov have suggested a
general expression for the expectation value of the vertex operator $\langle e^{ia\vartheta}\rangle$, [see Eq.(20) in that reference] 
of the sine-Gordon model given by the action
\begin{equation}
S_{sG}=\int d^2x\Big\{\frac{1}{16\pi}(\partial_\nu\vartheta)^2 - 2\mu\cos(\beta'\vartheta)\Big\}.
\label{eq:sgaction}
\end{equation}
Their conjecture is as follows (for $\beta'^2<1$, and $|{\rm Re} ~a|<1/2\beta'$, which are required for the convergence) ,
\begin{widetext}
	\begin{equation}
	\langle e^{ia\vartheta}\rangle=\Big[\frac{m\Gamma(\frac{1}{2}+\frac{\xi}{2})\Gamma(1-\frac{\xi}{2})}{4\sqrt{\pi}}\Big]^{2a^2}\exp{\Big\{\int_{0}^{\infty}\frac{dt}{t}\Big[\frac{\sinh^2(2a\beta' t)}{2\sinh(\beta'^2t)\sinh(t)\cosh\big((1-\beta'^2)t\big)}-2a^2e^{-2t}\Big]\Big\}},
	\label{eq:20}
	\end{equation}
\end{widetext}
where
\begin{equation}
m=2M\sin(\pi\xi/2),\quad \xi=\frac{\beta'^2}{1-\beta'^2}
\label{eq:m}
\end{equation}
with $M$ the soliton mass.

\subsection{Perturbation $H_C$ and $H_\sigma$}
\label{subsec:hcandhs}

Here we work out the action for our KT Hamiltonian by considering $H_C$ and $H_\sigma$ as perturbations to the harmonic Hamiltonian $H_0$.
Provided the field is small enough, so that the scaling dimensions of various operators are given by their values at the Heisenberg point,
we have 
\begin{equation}
\begin{split}
M_R^z&=\frac{1}{2\sqrt{2\pi}}(\partial_x\varphi-\partial_x\vartheta),\\ M_L^z&=\frac{1}{2\sqrt{2\pi}}(\partial_x\varphi+\partial_x\vartheta).
\end{split}
\end{equation}
and therefore 
\begin{equation}
\begin{split}
H_\sigma
&=-\frac{v y_\sigma}{4}\int \mathrm{d}x[(\partial_x\varphi)^2-(\partial_x\vartheta)^2],\\
H_C&=\frac{v y_C}{2\pi a^2}\int \mathrm{d}x \cos(2\sqrt{2\pi}\vartheta).
\end{split}
\end{equation}
Therefore, the action, which determines the partition function $Z = \int e^{-S}$, is
\begin{equation}
\begin{split}
S&=\int \mathrm{d}x\mathrm{d}\tau\Big\{-i\partial_x\vartheta\partial_\tau\varphi
+\frac{1}{2}[v_1(\partial_x\varphi)^2+v_2(\partial_x\vartheta)^2 ]\\
&\qquad\quad+\frac{v y_C}{2\pi a^2}\cos(\sqrt{8\pi}\vartheta)\Big\}.\\
\end{split}
\label{eq:SG2}
\end{equation}
where
\begin{equation}
v_1= v(1-\frac{y_\sigma}{2}),\quad v_2=v(1+\frac{y_\sigma}{2}).
\end{equation}
We integrate out the $\varphi$ field using the duality $\partial_x\vartheta\partial_\tau\varphi=\partial_x\varphi\partial_\tau\vartheta$ and then 
the action factorizes
\begin{eqnarray}
S=&&\int \mathrm{d}x\mathrm{d}\tau\Big\{\frac{v_1}{2}(\partial_x\varphi-\frac{i}{v_1}\partial_\tau\vartheta)^2+\frac{1}{2v_1}(\partial_\tau\vartheta)^2+\frac{v_2}{2}(\partial_x\vartheta)^2\nonumber\\
&&+\frac{v y_C}{2\pi a^2}\cos(\sqrt{8\pi}\vartheta)\Big\}.
\end{eqnarray}
The first, $\varphi$-dependent piece in Eq.~\eqref{eq:SG2} is integrated away.
The remaining $\vartheta$ part of the action is
\begin{eqnarray}
S_\vartheta&=&\int \mathrm{d}xdy\Big\{\frac{1}{2}\sqrt{\frac{v_2}{v_1}}((\partial_x\vartheta)^2+(\partial_\tau\vartheta)^2)\nonumber\\
&&+\frac{y_C}{2\pi a^2}\frac{v}{u}\cos(\sqrt{8\pi}\vartheta)\Big\},
\end{eqnarray}
with $y=u\tau$ and set $u=\sqrt{v_1v_2}$.
Finally, we rescale $\vartheta$,
\begin{equation}
\vartheta=\frac{1}{\sqrt{8\pi}}\Big(\frac{v_1}{v_2}\Big)^\frac{1}{4}\tilde{\vartheta}, 
\end{equation}
and arrive at the desired form of Eq.~\eqref{eq:sgaction},
\begin{equation}
S_\vartheta=\int d^2x\Big\{\frac{1}{16\pi}(\partial_\nu\tilde\vartheta)^2-2\mu\cos(\tilde{\beta}\tilde{\vartheta})\Big\},\\
\end{equation}
where
\begin{equation}
\mu\equiv\frac{|y_C|}{4\pi a^2}\frac{v}{u},\quad\tilde{\beta}\equiv\Big(\frac{v_1}{v_2}\Big)^\frac{1}{4}.
\label{eq:params1}
\end{equation}
Here, for the case of $y_C>0$, we made an additional shift $\tilde{\vartheta}\to \tilde{\vartheta}+\pi/\tilde{\beta}$ 
in order to change the sign of the cosine term.
The case of $y_C<0$ does not require any additional shifts, $\tilde{\vartheta}= \tilde{\vartheta}$.
The parameters \eqref{eq:params1} of the action can easily be written in terms of $y_{C,\sigma}$,
\begin{equation}
\begin{gathered}
u=v\sqrt{1-y_\sigma^2/4},\quad \mu=\frac{1}{4\pi a^2}\frac{{|y_C|}}{\sqrt{1-y_\sigma^2/4}},\\
\tilde{\beta}=\Big(\frac{1-y_\sigma/2}{1+y_\sigma/2}\Big)^\frac{1}{4}.
\label{eq:parameter2}
\end{gathered}
\end{equation}
The expectation value we intend to compute is $\langle e^{i\sqrt{2\pi}\vartheta}\rangle=\langle e^{i\tilde{\beta}\tilde{\vartheta}/2}\rangle$, 
and thus $a$ in Eq.~\eqref{eq:20} is just $a\equiv \tilde{\beta}/2$. 

We observe that our order parameters are obtained as $N^y \sim \cos\sqrt{2\pi}\vartheta \propto {\rm Re}\langle e^{i\tilde{\beta}\tilde{\vartheta}/2}\rangle$, while 
$N^z \sim \sin\sqrt{2\pi}\vartheta \propto {\rm Im}\langle e^{i\tilde{\beta}\tilde{\vartheta}/2}\rangle$.
The shift described just below Eq.~\eqref{eq:params1}, which is needed for $y_C>0$, transforms $\langle e^{i\tilde{\beta}\tilde{\vartheta}/2}\rangle$
into $e^{i \pi/2} \langle e^{i\tilde{\beta}\tilde{\vartheta}/2}\rangle$ and thus precisely corresponds to the change of the order from the $N^y$ kind (realized for $y_C < 0$) to 
the $N^z$ kind (realized for $y_C > 0$).

\subsection{Order parameter $\langle e^{i\frac{\tilde{\beta}}{2}\tilde{\vartheta}}\rangle$ }

We are interested in evaluating the expectation value
\begin{equation}
\langle e^{i\frac{\tilde{\beta}}{2}\tilde{\vartheta}}\rangle=A e^I,\quad A\equiv\Big[\frac{m\Gamma(\frac{1}{2}+\frac{\xi}{2})\Gamma(1-\frac{\xi}{2})}{4\sqrt{\pi}}\Big]^{\tilde{\beta}^2/2}.
\end{equation}
Here $I$ is obtained from Eq.~\eqref{eq:20} by setting $a=\tilde{\beta}/2$,
\begin{equation}
\begin{gathered}
I\equiv\int_{0}^{\infty}\frac{dt}{t}\Big[\frac{\sinh(\tilde{\beta}^2 t)}{2\sinh(t)\cosh\big((1-\tilde{\beta}^2)t\big)}-\frac{\tilde{\beta}^2}{2}e^{-2t}\Big].
\end{gathered}
\end{equation}
The convergence of $I$ is easy to check: $\tilde{\beta}^2<1$ is required for $t\to \infty$.
Using the identity $\Gamma(1-x)\Gamma(x)={\pi}/{\sin(\pi x)}$, and with $m$ in Eq.~\eqref{eq:m}, the expression for $A$ becomes
\begin{equation}
\begin{split}
A&=\Big[\frac{\sqrt{\pi}}{2}M\frac{\Gamma(\frac{1}{2}+\frac{\xi}{2})}{\Gamma(\xi/2)}\Big]^{\tilde{\beta}^2/2}.
\end{split}
\label{eq:m1}
\end{equation}
The relation between constant $\mu$ and mass $M$ is [this is Eq. (12) of Ref.~\onlinecite{Lukyanov1997}]
\begin{equation}
\mu=\frac{\Gamma(\tilde{\beta}^2)}{\pi \Gamma(1-\tilde{\beta}^2)}\Big[M\frac{\sqrt{\pi}\Gamma(\frac{1}{2}+\frac{\xi}{2})}{2\Gamma(\frac{\xi}{2})}\Big]^{2-2\tilde{\beta}^2}.
\end{equation}
Using all these we obtain for the order parameter
\begin{widetext}
	\begin{equation}
	\langle e^{i\frac{\tilde{\beta}}{2}\tilde{\vartheta}}\rangle=\Big[\frac{\pi\mu\;\Gamma(1-\tilde{\beta}^2)}{\Gamma(\tilde{\beta}^2)}\Big]^{\tilde{\beta}^2/[4(1-\tilde{\beta}^2)]}\times\exp{\Big\{ \int_{0}^{\infty}\frac{dt}{t}\Big[\frac{\sinh(\tilde{\beta}^2 t)}{2\sinh(t)\cosh\big((1-\tilde{\beta}^2)t\big)}-\frac{\tilde{\beta}^2}{2}e^{-2t}\Big]\Big\}}.
	\label{eq:expec}
	\end{equation}
\end{widetext}
Note that Eq.~\eqref{eq:expec} is a function of $\tilde{\beta}$, which, in turn, is function of running $y_\sigma(\ell)$. It also depends on running $y_C(\ell)$, 
via a $\mu$ dependence [see \eqref{eq:parameter2}].
Thus \eqref{eq:expec} allows us to evaluate the order parameter as a function of the RG scale $\ell$. 

\subsection{Luttinger liquid phase}
\label{app:LL}

The LL phase of our model is characterized by $y_C = 0 and y_\sigma < 0$ for $\ell \to \infty$ (see Fig.~\ref{fig:kt}). Correspondingly, its action is given by 
Eq.~\eqref{eq:SG2} with $y_C = 0$. From here it is easy to derive that the scaling dimension of the vertex operator $e^{i\sqrt{2\pi} \vartheta(x)}$ is 
$\Delta_\vartheta = {\tilde\beta}^2/2 \approx  (1 -{y_\sigma}/{2})/2$, while that of the dual field one $e^{i\sqrt{2\pi} \varphi(x)}$ is given by 
$\Delta_\varphi =1/(2 \tilde\beta^2) \approx (1 + {y_\sigma}/{2})/2$. Backscattering  renormalizes scaling dimensions through the RG flow of $y_\sigma$. Given that in the LL $y_\sigma < 0$, 
we observe that $\Delta_\varphi < \Delta_\vartheta$ which signals that the correlation functions of fields ${\cal N}^z$ and $\xi$, which are written
in terms of $\varphi$ bosons, decay {\em slower} than those of fields ${\cal N}^x$ and ${\cal N}^y$, which are expressed via $\vartheta$ bosons.
Moreover, due to Eq.~\eqref{shift:J-N}, correlations of ${\cal N}^z$ and $\xi$ are incommensurate,
\begin{equation}
\langle {\cal N}^z(x) {\cal N}^z(0) \rangle \propto \langle \xi(x) \xi(0) \rangle \propto \frac{\cos[t_\varphi x]}{|x|^{2 \Delta_\varphi}}
\label{eq:LL2}
\end{equation}
while those of ${\cal N}^{x,y}$ are commensurate
\begin{equation}
\langle {\cal N}^{x,y}(x) {\cal N}^{x,y}(0) \rangle \propto \frac{1}{|x|^{2 \Delta_\vartheta}} .
\label{eq:LL3}
\end{equation}
Taken together with Eq.~\eqref{eq:Neps}, which describes the relation between spin operators in the laboratory and rotated frames, these simple relations allow
us to fully describe the asymptotic spin (and dimerization) correlations in the LL phase with fully broken spin-rotational symmetry
\begin{eqnarray}
&&\langle S^x(x) S^x(0) \rangle \propto \frac{\cos[(\pi-t_\varphi )x]}{|x|^{2 \Delta_\varphi}},\nonumber\\
&&\langle S^y(x) S^y(0) \rangle \propto \sin^2\theta_0 \frac{\cos[(\pi-t_\varphi )x]}{|x|^{2 \Delta_\varphi}} + 
\cos^2\theta_0 \frac{(-1)^x}{|x|^{2 \Delta_\vartheta}},\nonumber\\
&&\langle S^z(x) S^z(0) \rangle \propto \frac{(-1)^x}{|x|^{2 \Delta_\vartheta}},\\
&& \langle \epsilon(x) \epsilon(0) \rangle \propto  \cos^2\theta_0 \frac{\cos[(\pi-t_\varphi )x]}{|x|^{2 \Delta_\varphi}} + 
\sin^2\theta_0 \frac{(-1)^x}{|x|^{2 \Delta_\vartheta}} .\nonumber
\label{eq:LL4}
\end{eqnarray}
Due to $\Delta_\varphi < \Delta_\vartheta$, the LL phase is dominated by the incommensurate correlations of $S^{x,y}$ and $\epsilon$ fields. Their contribution to the equal time structure factor is easy to estimate by simple scaling analysis. For example, defining $Q = \pi - t_\varphi$, we have
\begin{equation}
M_s^x(k) \propto \int dx \frac{e^{i (k-Q) x}}{|x|^{2 \Delta_\varphi}} \sim |k-Q|^{{2 \Delta_\varphi} - 1},
\end{equation}
where we extended the limits of the integration to infinity due to convergence of the integral for $2 \Delta_\varphi > 0$. The divergence at $k=Q$ is 
controlled by $2 \Delta_\varphi - 1 = - y_\sigma/2 < 0$
and is rounded in the system of finite size $L$. More careful calculation of $M_s^a(k)$ and $M_d(k)$ is possible \cite{Hikihara1998,Hikihara2004,Hikihara2017}, but is beyond the scope of the present study.

%%%%%%%%%%%%%%%%%%%%%%%%%%%%%%%
\section{DMRG details} \label{app:DMRG}%

\begin{figure}[htb]
   \centering
     \includegraphics[width=0.85\linewidth]{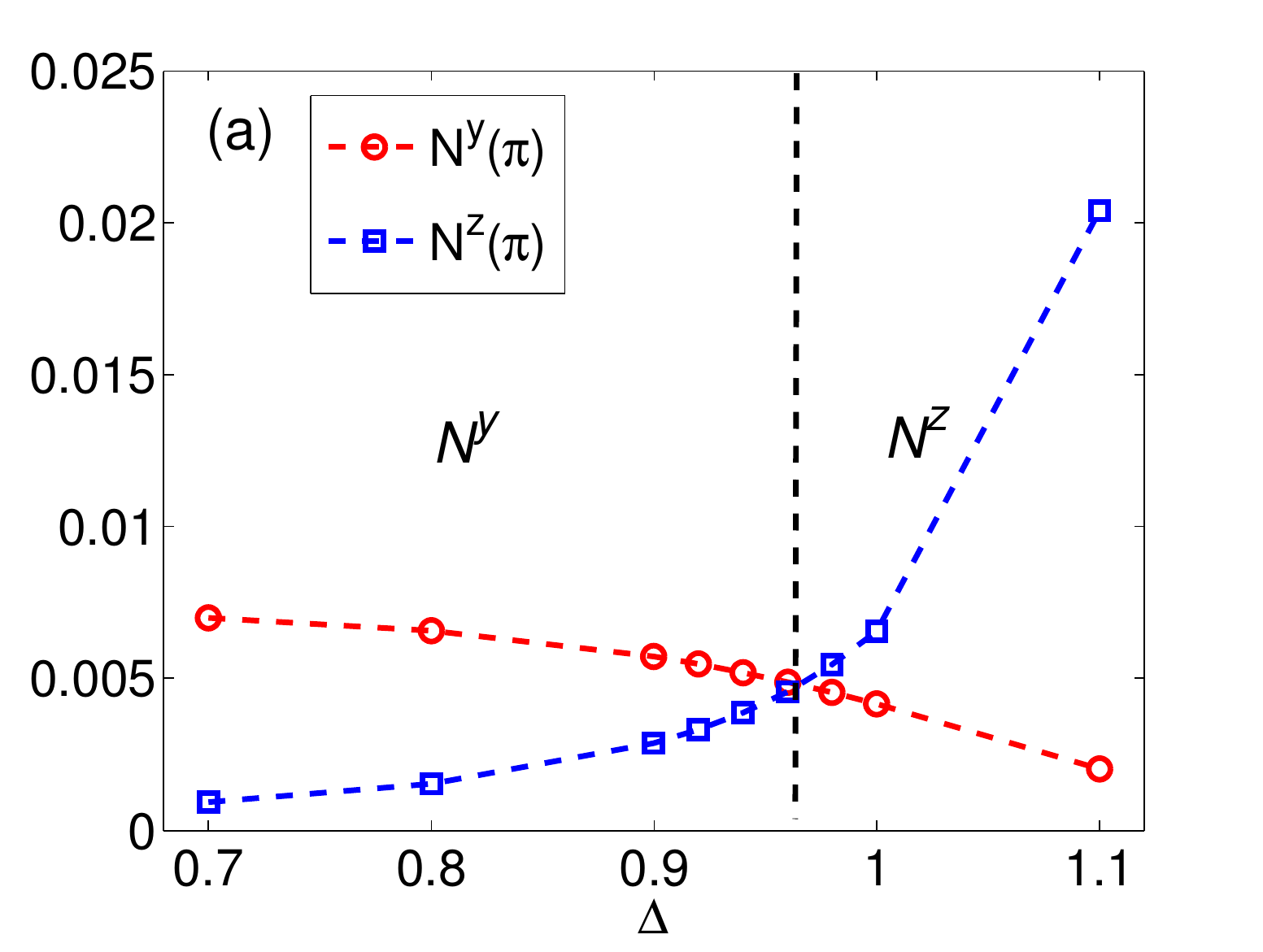}
     \includegraphics[width=0.85\linewidth]{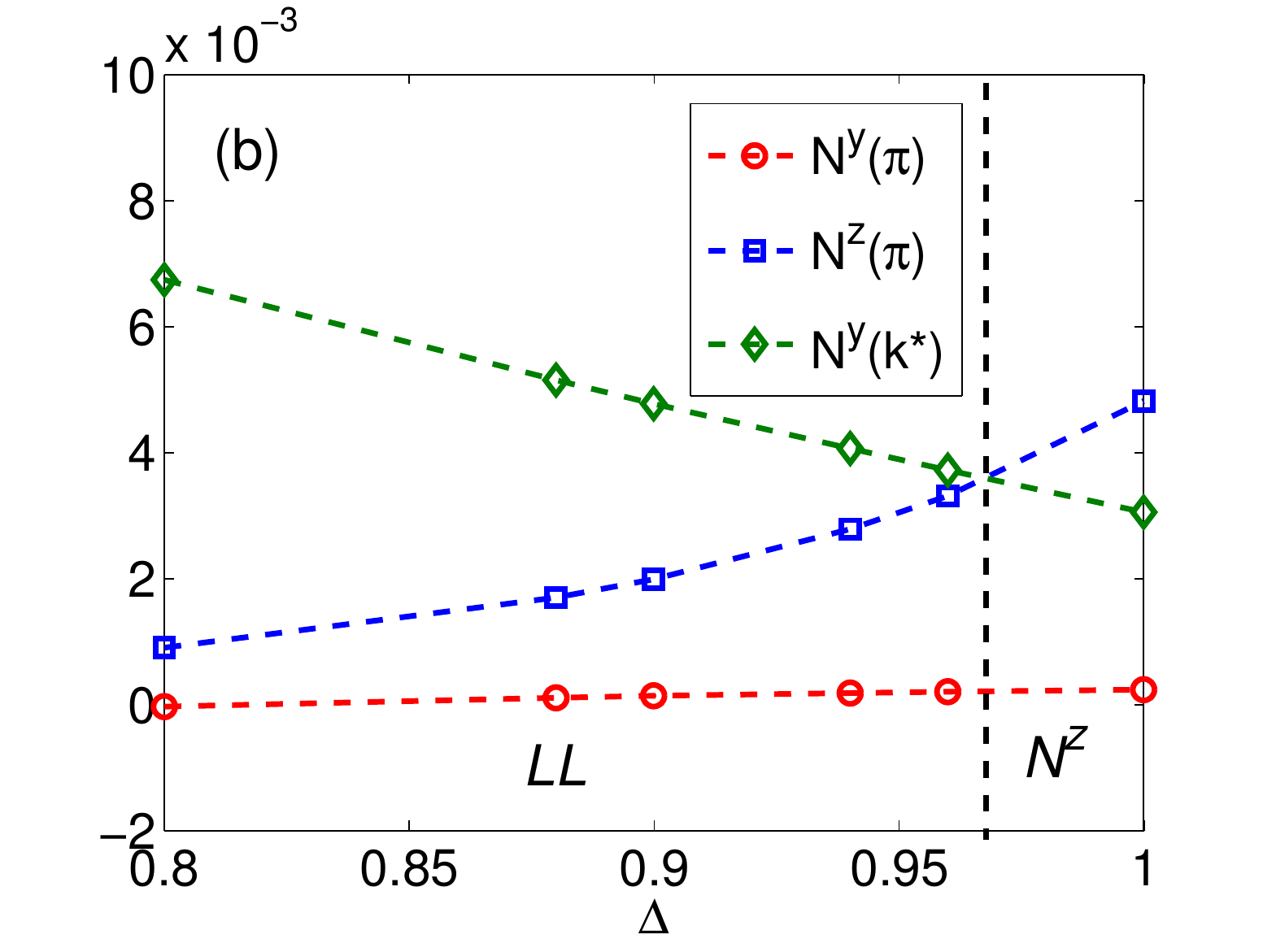}
\caption{(Color online) Order parameters $N^y(\pi)$ (red circles), $N^z(\pi)$ (blue squares) and $N^y(k^*)$ (green diamonds) extrapolated by a second order polynomial \eqref{eq:fit}
using data from $L=600$, $800$, $1000$, $1200$, and $1600$ chains as a function of $\Delta$ at (a) $h/J=0.2$ and (b) $h/J=0.05$ with $D/J=0.1$. 
The crossing points of the order parameters determine the phase boundary.}
\label{fig:order}
\end{figure}

In this appendix, we provide details on the determination of the phase diagram and finite-size effects.

\subsection{Determination of phase boundaries}%%%%%%%%%%%

\begin{figure}[htb]
\begin{center}
\includegraphics[width=0.9\linewidth]{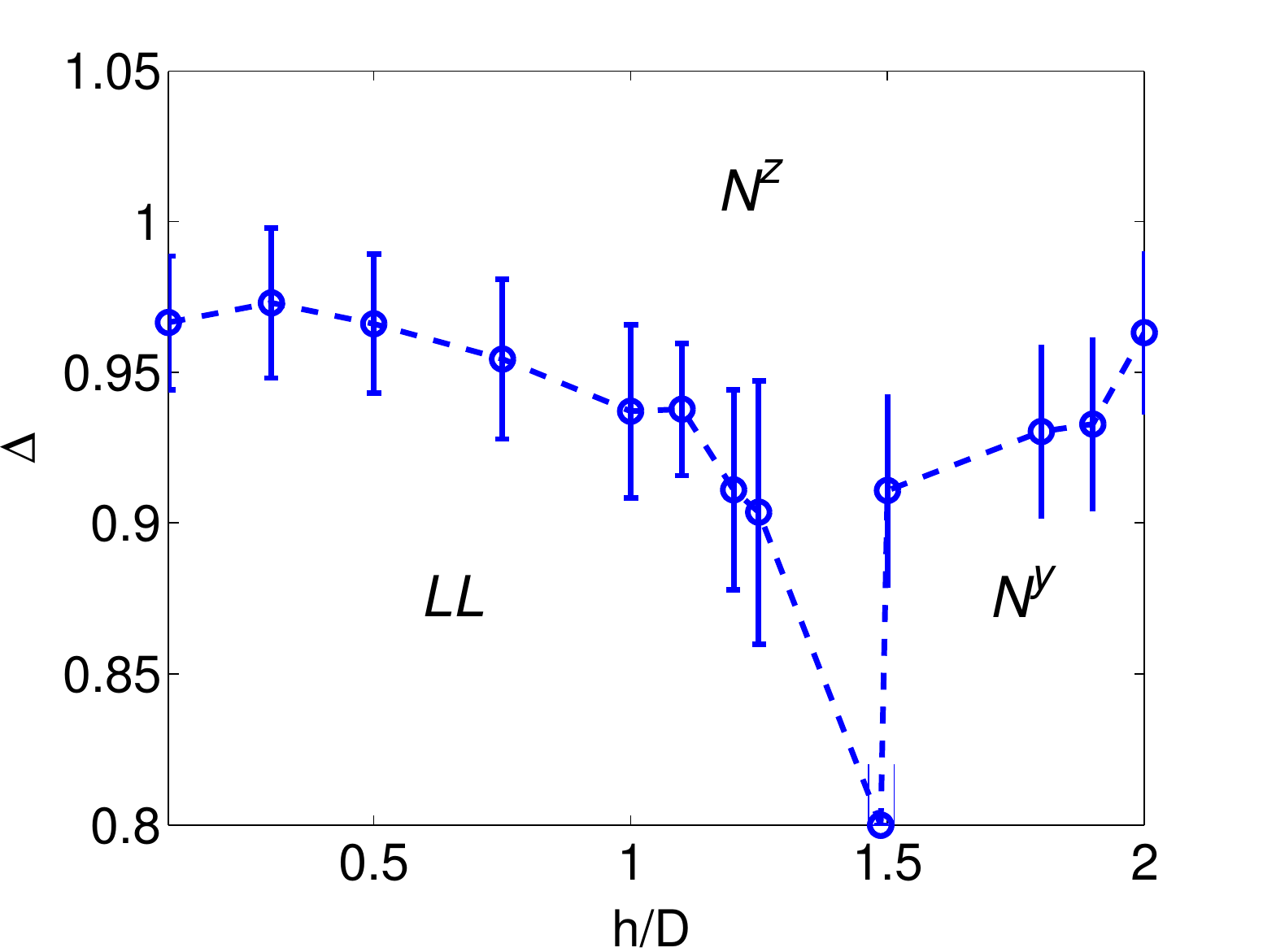}
\caption{(Color online) Phase diagram of the chain with $D/J=0.1$ after the finite-size extrapolation of the order parameters to $L=\infty$ using Eq.\eqref{eq:fit}.
The error bar are plotted at the 95\% confidence interval of the order parameters.} 
\label{fig:extrap_phase}
\end{center}
\end{figure}

Here we describe how we determine phase boundaries numerically. In Fig.~\ref{fig:order}(a) we show the extrapolated order parameters at $k=\pi$ near the phase boundary between the 
$N^y$ and $N^z$ phases. Here we can see that these two distinct orders are dominant in the corresponding phases, hence the phase boundary between them can be determined by their crossing point. 

Figure~\ref{fig:order}(b) shows the order parameters near the boundary between the LL and $N^z$ phases, 
where both order parameters $N^y(k=k^*)$ and $N^z(k=\pi)$ are finite and dominant on the opposite sides of the figure, while $N^y(k=\pi)$ is vanishingly small. Notice that due to large finite-size effect, the order parameter $N^y(k=k^*)$, which should vanish after extrapolation to the thermodynamic limit
$L\to \infty$, still remains finite in our $L = 1600$ chain, although rather small.
As a result, we use it to identify the LL phase as described in the main text.

\subsection{Finite size effects on the phase boundary}

To check the finite-size effect on phase boundaries,
we have compared phase diagrams for the chain of length $L=1200$ calculated by DMRG and iTEBD methods as shown in Fig.~\ref{fig:phase}. 
To minimize the boundary effect, the order parameters are calculated within the central half of the system, i.e., 600 sites in the middle of the system. 
We keep the same bond-link dimension and considering the same lengths for the calculation of correlation functions using 
iTEBD and DMRG methods. The agreement between the DMRG and iTEBD results is quite good, suggesting that the DMRG results are only subject to the finite size effect while the effect of open boundaries is negligible. 

Figure~\ref{fig:extrap_phase} shows the phase diagram obtained by extrapolating order parameters to $L=\infty$ using second-order polynomial functions of $1/\sqrt{L}$ [Eq.~\eqref{eq:fit}].
Comparing it to the phase diagram in Fig.~\ref{fig:phase} for the finite system of size $L=1200$, we observe the shift of the $N^z$-$LL$ and $N^z$-$N^y$ boundaries to slightly larger $\Delta$ values.
A more detailed analysis suggests that error bars associated with the finite-size extrapolation to $L=\infty$ are within a 95\% confidence interval, 
which means that our conclusion about the $N^z$ Ising order extending to the $\Delta<1$ region is well justified.

It is also possible to determine the phase boundary by computing the Binder cumulant\cite{Pelissetto2002,West2015,Saadatmand2015}, which is widely used in Monte Carlo studies and has also been recently applied in the DMRG study \cite{West2015,Saadatmand2015}. Our preliminary investigation suggests that the phase boundary determined with the help of the Binder cumulant is fully consistent with the results obtained in this work.

\bibliography{Refs}

\end{document}